# CUORE-0 detector: design, construction and operation

---

## CUORE Collaboration

**C. Alduino,[a] K. Alfonso,[b] D. R. Artusa,[a,c] F. T. Avignone III,[a] O. Azzolini,[d] M. Balata,[c] T. I. Banks,[e,f] G. Bari,[g] J.W. Beeman,[h] F. Bellini,[i,j] A. Bersani,[k] D. Biare,[f] M. Biassoni,[l,m] F. Bragazzi,[k] C. Brofferio,[l,m] A. Buccheri,[j] C. Bucci,[c] C. Bulfon,[j] A. Caminata,[k] L. Canonica,[c,u] X. G. Cao,[n] S. Capelli,[l,m] M. Capodiferro,[j] L. Cappelli,[k,c,o] L. Carbone,[m] L. Cardani,[i,j,1] M. Cariello,[k] P. Carniti,[l,m] N. Casali,[i,j] L. Cassina,[l,m] R. Cereseto,[k] G. Ceruti,[m] A. Chiarini,[g] D. Chiesa,[l,m] N. Chott,[a] M. Clemenza,[l,m] D. Conventi[d] S. Copello,[p,k] C. Cosmelli,[i,j] O. Cremonesi,[m,2] R. J. Creswick,[a] J. S. Cushman,[q] A. D'Addabbo,[c] I. Dafinei,[j] C. J. Davis,[q] S. Dell'Oro,[c,r] M. M. Deninno,[g] S. Di Domizio,[p,k] M. L. Di Vacri,[c,s] L. DiPaolo,[f] A. Drobizhev,[e,f] G. Erme,[c,o] D. Q. Fang,[n] M. Faverzani,[l,m] G. Fernandes,[p,k] E. Ferri,[l,m] F. Ferroni,[i,j] E. Fiorini,[m,l] S. J. Freedman,[f,e,3] B. K. Fujikawa,[f] R. Gaigher,[m] A. Giachero,[m] L. Gironi,[l,m] A. Giuliani,[t] L. Gladstone,[u] P. Gorla,[c] C. Gotti,[l,m] M. Guetti,[c] T. D. Gutierrez,[v] E. E. Haller,[h,w] K. Han,[x,q] E. Hansen,[u,b] K. M. Heeger,[q] R. Hennings-Yeomans,[e,f] K. P. Hickerson,[b] H. Z. Huang,[b] M. Iannone,[j] L. Ioannucci,[c] R. Kadel,[y] G. Keppel,[d] Yu. G. Kolomensky,[e,y,f] A. Leder,[u] K. E. Lim,[q] X. Liu,[b] Y. G. Ma,[n] M. Maino,[l,m] L. Marini,[p,k] M. Martinez,[i,j,z] R. H. Maruyama,[q] R. Mazza,[m] Y. Mei,[f] S. Meijer,[v,f] R. Michinelli,[g] D. Miller,[v,f] N. Moggi,[aa,g] S. Morganti,[j] P. J. Mosteiro,[j] M. Nastasi,[l,m] S. Nisi,[c] C. Nones,[ab] E. B. Norman,[ac,ad] A. Nucciotti,[l,m] T. O'Donnell,[e,f] F. Orio,[j] D. Orlandi,[c] J. L. Ouellet,[u,e,f] C. E. Pagliarone,[c,o] M. Pallavicini,[p,k] V. Palmieri,[d] G. Pancaldi,[g] L. Pattavina,[c] M. Pavan,[l,m] R. Pedrotta,[ae] A. Pelosi,[j] M. Perego,[m] G. Pessina,[m] V. Pettinacci,[j] G. Piperno,[i,j] S. Pirro,[c] S. Pozzi,[l,m] E. Previtali,[m] C. Rosenfeld,[a] C. Rusconi,[m] E. Sala,[l,m] S. Sangiorgio,[ac] D. Santone,[c,s] N. D. Scielzo,[ac] V. Singh,[e] M. Sisti,[l,m] A. R. Smith,[f] F. Stivanello,[d] L. Taffarello,[ae] L. Tatananni,[c] M. Tenconi,[t] F. Terranova,[l,m] M. Tessaro,[ae] C. Tomei,[j] S. Trentalange,[b] G. Ventura,[af,ag] M. Vignati,[j] S. L. Wagaarachchi,[e,f] J. Wallig,[ah] B. S. Wang,[ac,ad] H. W. Wang,[n] J. Wilson,[a] L. A. Winslow,[u] T. Wise,[q,ai] L. Zanotti,[l,m] C. Zarra,[c] G. Q. Zhang,[n] B. X. Zhu,[b] S. Zimmermann,[ah] and S. Zucchelli[aj,g]**

[a]*Department of Physics and Astronomy, University of South Carolina, Columbia, SC 29208 - USA*
[b]*Department of Physics and Astronomy, University of California, Los Angeles, CA 90095 - USA*
[c]*INFN - Laboratori Nazionali del Gran Sasso, Assergi (L'Aquila) I-67010 - Italy*
[d]*INFN - Laboratori Nazionali di Legnaro, Legnaro (Padova) I-35020 - Italy*
[e]*Department of Physics, University of California, Berkeley, CA 94720 - USA*

---

[1]Present address: Physics Department, Princeton University, Princeton, NJ 08544, USA.
[2]Corresponding author.
[3]Deceased.
[4]Present address: INFN Laboratori Nazionali di Frascati, Via E. Fermi, 40 – I-00044 Frascati (Rome), Italy.

[f] Nuclear Science Division, Lawrence Berkeley National Laboratory, Berkeley, CA 94720 - USA
[g] INFN - Sezione di Bologna, Bologna I-40127 - Italy
[h] Materials Science Division, Lawrence Berkeley National Laboratory, Berkeley, CA 94720 - USA
[i] Dipartimento di Fisica, Sapienza Università di Roma, Roma I-00185 - Italy
[j] INFN - Sezione di Roma, Roma I-00185 - Italy
[k] INFN - Sezione di Genova, Genova I-16146 - Italy
[l] Dipartimento di Fisica, Università di Milano-Bicocca, Milano I-20126 - Italy
[m] INFN - Sezione di Milano Bicocca, Milano I-20126 - Italy
[n] Shanghai Institute of Applied Physics, Chinese Academy of Sciences, Shanghai 201800 - China
[o] Dipartimento di Ingegneria Civile e Meccanica, Università degli Studi di Cassino e del Lazio Meridionale, Cassino I-03043 - Italy
[p] Dipartimento di Fisica, Università di Genova, Genova I-16146 - Italy
[q] Department of Physics, Yale University, New Haven, CT 06520 - USA
[r] INFN - Gran Sasso Science Institute, L'Aquila I-67100 - Italy
[s] Dipartimento di Scienze Fisiche e Chimiche, Università dell'Aquila, L'Aquila I-67100 - Italy
[t] CSNSM, Univ. Paris-Sud, CNRS/IN2P3, Université Paris-Saclay, 91405 Orsay, France
[u] Massachusetts Institute of Technology, Cambridge, MA 02139 - USA
[v] Physics Department, California Polytechnic State University, San Luis Obispo, CA 93407 - USA
[w] Department of Materials Science and Engineering, University of California, Berkeley, CA 94720 - USA
[x] Department of Physics and Astronomy, Shanghai Jiao Tong University, Shanghai 200240 - China
[y] Physics Division, Lawrence Berkeley National Laboratory, Berkeley, CA 94720 - USA
[z] Laboratorio de Fisica Nuclear y Astroparticulas, Universidad de Zaragoza, Zaragoza 50009 - Spain
[aa] Dipartimento di Scienze per la Qualità della Vita, Alma Mater Studiorum - Università di Bologna, Bologna I-47921 - Italy
[ab] Service de Physique des Particules, CEA / Saclay, 91191 Gif-sur-Yvette - France
[ac] Lawrence Livermore National Laboratory, Livermore, CA 94550 - USA
[ad] Department of Nuclear Engineering, University of California, Berkeley, CA 94720 - USA
[ae] INFN - Sezione di Padova, Padova I-35131 - Italy
[af] Dipartimento di Fisica, Università di Firenze, Firenze I-50125 - Italy
[ag] INFN - Sezione di Firenze, Firenze I-50125 - Italy
[ah] Engineering Division, Lawrence Berkeley National Laboratory, Berkeley, CA 94720 - USA
[ai] Department of Physics, University of Wisconsin, Madison, WI 53706 - USA
[aj] Dipartimento di Fisica e Astronomia, Alma Mater Studiorum - Università di Bologna, Bologna I-40127 - Italy

*E-mail:* cuore-spokesperson@lngs.infn.it

Abstract: The CUORE experiment will search for neutrinoless double-beta decay of $^{130}Te$ with an array of 988 $TeO_2$ bolometers arranged in 19 towers. CUORE-0, the first tower assembled according to the CUORE procedures, was built and commissioned at Laboratori Nazionali del Gran Sasso, and took data from March 2013 to March 2015. In this paper we describe the design, construction and operation of the CUORE-0 experiment, with an emphasis on the improvements made over a predecessor experiment, Cuoricino. In particular, we demonstrate with CUORE-0 data that the design goals of CUORE are within reach.

Keywords: Double-beta decay detectors; Cryogenic detectors; Cryogenics; Detector control systems (detector and experiment monitoring and slow-control systems, architecture, hardware, algorithms, databases)

## Contents

## 1 Introduction

Neutrinoless double-beta ($0\nu\beta\beta$) decay [1] is an extremely rare process, if it occurs at all, in which a nucleus undergoes two simultaneous beta decays without emitting any neutrinos. This can occur only if neutrinos have nonzero mass and are "Majorana particles" [2]; that is, if they are their own anti-particles. In recent years, neutrino oscillation experiments have conclusively established that neutrinos do indeed have mass, but these experiments are only sensitive to the differences between the squares of those masses and not the absolute masses themselves [3]. The search for $0\nu\beta\beta$ decay is therefore important for several reasons: it is a theoretically well motivated and experimentally feasible mode in which to discover lepton number violation; it can establish if the neutrino is a Majorana particle, and thus unique among the known constituents of matter; and, if the process is observed, it can constrain the absolute neutrino mass scale. To date there have been no conclusive observations of $0\nu\beta\beta$ decay, but recent results have placed very stringent upper limits on the decay rate in $^{76}$Ge [4], $^{130}$Te [5], $^{136}$Xe [6, 7] and $^{100}$Mo [8].

The Cryogenic Underground Observatory for Rare Events (CUORE) [9–11], is an upcoming cryogenic bolometric experiment designed to search for $0\nu\beta\beta$ decay in $^{130}$Te. The CUORE detector will consist of a close-packed array of 988 $TeO_2$ bolometers. This corresponds to 206 kg of $^{130}$Te, considering the natural abundance of 34.2%. The array will be cooled inside a large cryostat to 10 mK; at this low temperature the crystals function as calorimeters, converting the minute energies deposited by particles inside them into measurable rises in temperature.

$TeO_2$ bolometers have long been used to search for $0\nu\beta\beta$ decay [12–15] because they satisfy many of the key constraints necessary to mount a sensitive experiment. These bolometers achieve high detection efficiency since they are both the source and detector of $0\nu\beta\beta$ decay; crystals that have low intrinsic background and exhibit excellent energy resolution can be grown reproducibly; increasingly large arrays have been operated with progressively improving stability for extended periods (i.e., several years). Furthermore, of the nuclei appropriate for double-beta decay searches, $^{130}$Te has the highest natural isotopic abundance and a relatively high endpoint energy ($Q_{\beta\beta}$) of $2527.518 \pm 0.013$ keV [16–18]. Finally $TeO_2$ satisfies the mechanical and cryogenic constraints for a bolometric experiment.

Until recently, the best limit on $0\nu\beta\beta$ decay in $^{130}$Te, $T_{1/2}^{0\nu\beta\beta} > 2.8 \times 10^{24}$ y (90% C.L.), came from Cuoricino [15]. The Cuoricino detector was a tower of 62 $TeO_2$ crystals, containing 11 kg of $^{130}$Te, and operated from 2003 to 2008 in a cryostat at the underground Laboratori Nazionali del Gran Sasso (LNGS), in Assergi, Italy. CUORE takes advantage of the experience from Cuoricino, but aims to improve the $0\nu\beta\beta$ decay half-life sensitivity by roughly two orders of magnitude. The CUORE bolometers will be arranged in a cylindrical matrix of 19 towers, each containing 13 planes of four crystals supported inside a copper frame. As a first step towards CUORE, before the construction of the 19 towers, we assembled a single tower (CUORE-0) and operated it in the former Cuoricino cryostat at LNGS. CUORE-0 served as a test of the new CUORE assembly line, both in terms of detector performance and control of radioactive background. Combining the exposure of CUORE-0 with Cuoricino yields the most stringent limit to date for the half-life of $0\nu\beta\beta$ decay of $^{130}$Te, $T_{1/2}^{0\nu\beta\beta} > 4.0 \times 10^{24}$ y (90% C.L.) [5].

In the following sections, we describe the CUORE-0 detector and its performance in detail. The design of the new $TeO_2$ bolometers is presented in Section 2, followed by a description of the CUORE-0 detector components in Section 3. After the description of the assembly technique in Section 4, an overview of the experimental setup is presented in Section 5. In Section 6 we describe the electronic readout and data acquisition. Section 7 is dedicated to the CUORE-0 results on bolometer reproducibility and uniformity, energy resolution, and background rate.

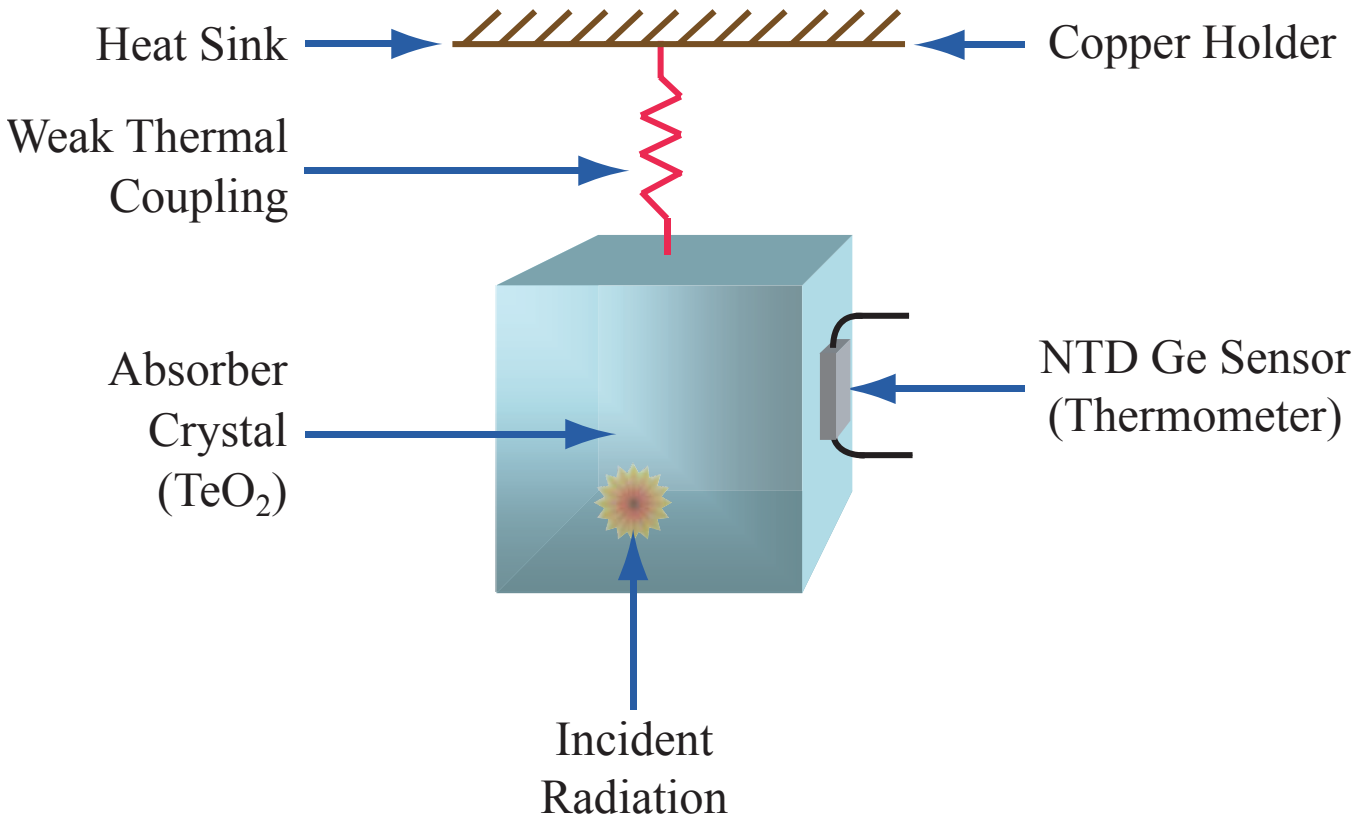

**Figure 1**. Schematic of a bolometric detector. A CUORE bolometer consists of an absorber connected to a heat sink through a weak thermal link, and read out by a temperature sensor attached to the absorber.

## 2 Detector overview

### 2.1 The bolometers tower

A bolometric detector is an extremely sensitive low-temperature calorimeter [19]. It consists of three main components: an energy absorber, a temperature sensor, and a thermal link to a heat sink (see Figure 1). The absorber is the target in which particles interact and deposit energy. The primary constraint on the absorber material is that its heat capacity must be small at low temperatures. CUORE and CUORE-0 use 750 g $TeO_2$ crystals as absorbers. At 10 mK their heat capacity is ~2 nJ/K, which results in a temperature rise of ~0.1 mK for an energy deposit of 1 MeV. The small temperature variations of the crystals are measured by a thermal sensor. CUORE-0 and CUORE use neutron-transmutation-doped (NTD) germanium thermistors [20], which are highly sensitive over a large temperature range.

CUORE-0 is an array of 52 $TeO_2$ bolometers, arranged into 13 planes, or "floors." Each floor has four cubic $5 \times 5 \times 5$ cm$^3$ crystals (see Figure 2). Two chips are glued to each crystal: an NTD thermistor to measure the temperature change (see Section 3.3) and a silicon resistor which acts as a Joule heater to inject power (see Section 3.4). The crystals are housed in a copper structure and kept in position by polytetrafluoroethylene (PTFE) holders; these, together with the glue used to couple the NTDs and heaters, are the only components in direct contact with the crystals. The PTFE supports are designed such that at low temperatures they hold the crystals firmly and compensate for the differential thermal contraction of copper and $TeO_2$. Together with the NTD thermistor and heater gold electrical connections, the PTFE serves as weak thermal links to the copper structure, which acts as a heat sink [21, 22].

The electrical signals from the bolometers are carried to the top of the tower by two sets of cables running on opposite sides of the tower. The sensors are directly bonded onto pads at one end of the cable, while the other end of the cable plugs into a connector that reads out the detector to the room temperature front-end electronics.

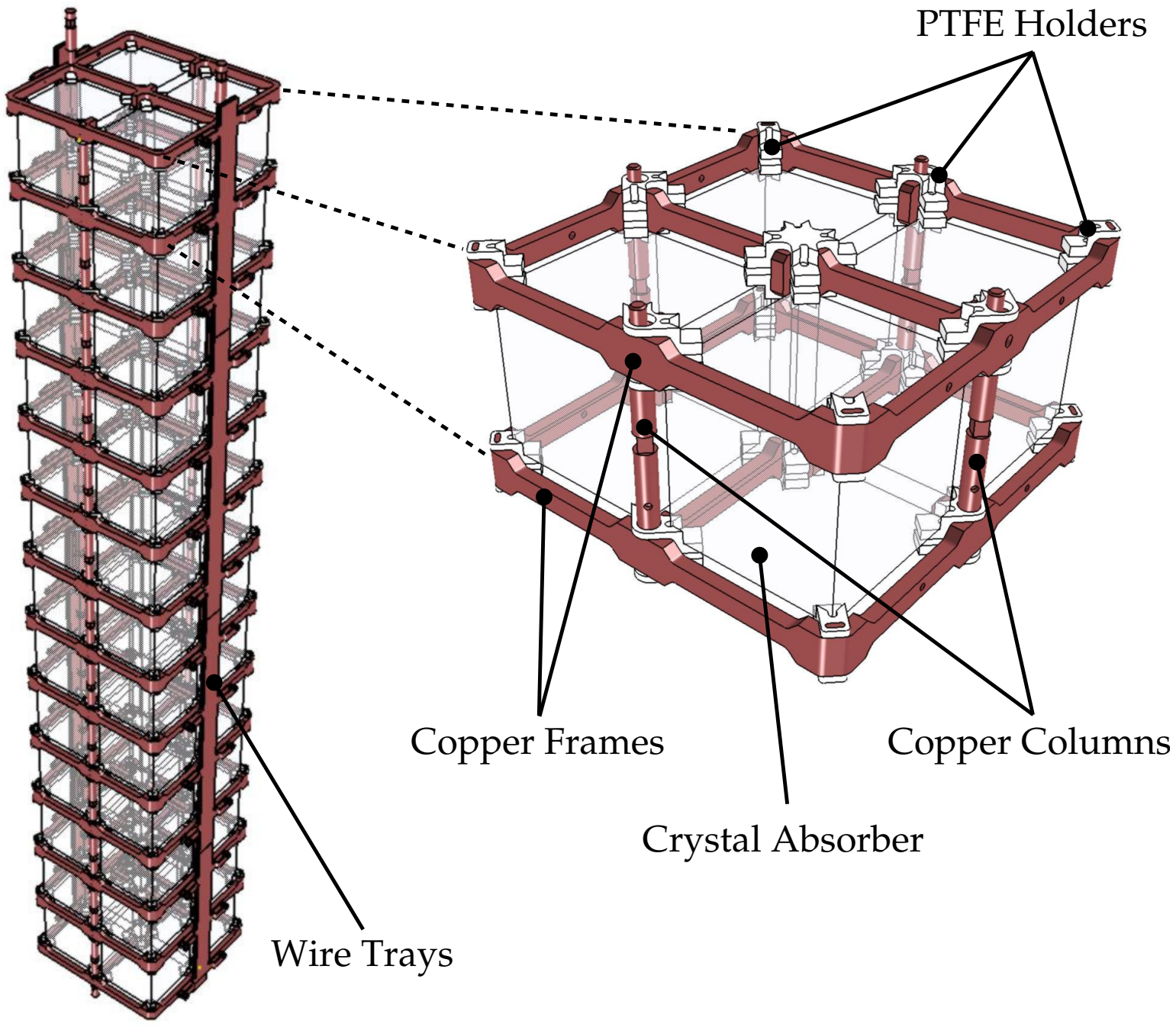

**Figure 2**. Left: A rendering of the CUORE-0 tower, including the wire trays for signal connection. Right: A single CUORE-0 detector floor, in which the $TeO_2$ crystals, PTFE holders, and copper frames and columns are indicated. The copper frames are shared between adjacent floors.

## 2.2 Guidelines for the design of the detector

The CUORE and CUORE-0 detector design was driven mainly by radioactivity and cryogenic considerations, which placed strong constraints on both the choice and amount of materials used. Bolometers are sensitive over their entire volume, so it is vital to minimize all possible background sources, including those capable of depositing energy at the surface of the crystals. The number and design of the individual components were optimized to fit the constraints posed by the machining and cleaning procedures and to assure a fast, reproducible and contaminant-free assembly process.

The detector design is the outcome of the experience acquired in running Cuoricino and other $TeO_2$ detector prototypes. To match mechanical, thermal and radioactivity requirements, the passive detector construction materials are mainly copper and PTFE. In Cuoricino, 70% of the overall background in the $0\nu\beta\beta$ decay region of interest (ROI) came from surface contamination of the $TeO_2$ crystals and materials facing them (mainly copper), particularly alpha particles produced by the $^{232}$Th and $^{238}$U primordial decay chains [23]. The remaining 30% of the background was caused by high energy beta and gamma interactions in the absorber, mainly from $^{232}$Th contamination in the construction materials of the cryostat. The background from cosmic-ray-induced particles was negligible for both Cuoricino and CUORE-0 [24].

To preselect materials with very low radioactive contamination for both the sensitive and passive parts of the detector all materials were scrutinized using a variety of radio-assay tech-niques. Alpha and gamma spectroscopy and Inductively Coupled Plasma Mass Spectrometry (ICPMS) analysis were performed at LNGS [25], Milano-Bicocca [26], Baradello laboratory [27] and Lawrence Berkeley National Laboratory (LBNL) [28]. Neutron activation analysis was carried out in collaboration with the Laboratory of Applied Nuclear Energy (LENA) in Pavia, Italy [29, 30].

To further reduce backgrounds from detector materials, we adopted two strategies: (1) minimize the copper mass close to the crystals and (2) develop procedures to avoid introduction of contaminants during the construction and storage of the detector components.

The first strategy impacts the detector design. In Cuoricino, each floor of the tower was assembled as a separate module, consisting of four $TeO_2$ crystals and two copper frames acting as crystal holders (above and below), and all the modules were stacked on top of each other. In CUORE-0, the tower was assembled as a single unit, with each copper frame acting as both the top and bottom holder of adjacent floors, as shown in Figure 2.

This new design reduces the total amount of copper interposed between the crystal planes by a factor of 2.3 in mass and 1.8 in surface area. This is beneficial as the copper has been identified as a major contributor to the background rate in the ROI. Moreover, the new design results in a detector structure that is more compact, with the crystals closer to one other and with less passive material in between. This allows for an improved tagging efficiency of background events from crystal surface contamination, as the particles emitted by the contaminants are more likely to deposit energy in neighbouring bolometers simultaneously and can be vetoed more efficiently with a coincidence analysis.

During production, we adopted new techniques for the crystal surface polishing and new cleaning procedures for the copper support structure. These new procedures are described in detail in Section 3.1 and Section 3.2, respectively. To prevent recontamination after the components were polished and cleaned, the tower assembly was carried out in custom-designed glove boxes [31] flushed with nitrogen gas. Assembly in a nitrogen atmosphere prevents $^{222}$Rn-induced surface contamination, which can be one of the most insidious backgrounds in low-background experiments [32]. To prevent recontamination of sensitive detector parts during long-term storage prior to being used, new packaging techniques, described in [33], were developed.

## 3 Detector components

For each component used in the detector specific selection processes, purification techniques, and assays were adopted to ensure cleanliness and radiopurity. We describe the design and results on the radiopurity of each component below; a summary is given in Table 1.

### 3.1 $TeO_2$ crystals

The $TeO_2$ crystals constitute the largest fraction of the detector mass. CUORE-0 crystals have a cubic shape with 5-cm sides and a mass of 750 g (see Figure 3). They have dimensional tolerance of 0.1% and extremely low levels of radioactive impurities. The $TeO_2$ crystals for CUORE-0 were produced by the Shanghai Institute of Ceramics, Chinese Academy of Sciences (SICCAS) in Shanghai, China. The raw tellurium metal and the $TeO_2$ powder were assayed using high-purity germanium detectors and ICPMS, which showed contamination levels below $2 \times 10^{-10}$ g/g (90% C.L.) in $^{232}$Th and $^{238}$U. The crystals were grown using the Bridgman process in platinum crucibles. Two successive crystal growths, starting from high-purity synthesized $TeO_2$ powder, were key to obtaining high-quality crystals both in terms of radiopurity and bolometric performance as described in [34]. After growth, the raw crystal ingots were subjected to rough mechanical processing (cutting, orienting, and shaping) followed by final surface polishing and packaging. All of these operations were performed in a dedicated clean room at the producer site. Strict controls were adopted at each stage of the crystal production to limit bulk and surface contaminations.

**Table 1**. Upper limits (at 90% C.L.) for bulk radioactive contaminations of the detector components.

| Component | $^{232}$Th [g/g] | $^{238}$U [g/g] |
|---|---|---|
| $TeO_2$ crystals | $< 2.1 \times 10^{-13}$ | $< 5.3 \times 10^{-14}$ |
| NOSV copper | $< 5.0 \times 10^{-13}$ | $< 5.3 \times 10^{-12}$ |
| NTD sensors | $< 1.0 \times 10^{-9}$ | $< 1.0 \times 10^{-9}$ |
| Bonding gold wires | $< 1.0 \times 10^{-8}$ | $< 1.0 \times 10^{-9}$ |
| Si heaters | $< 8 \times 10^{-11}$ | $< 1.7 \times 10^{-10}$ |
| PTFE holders | $< 1.5 \times 10^{-12}$ | $< 1.8 \times 10^{-12}$ |
| Cu-PEN cables | $< 4.4 \times 10^{-10}$ | $< 1.1 \times 10^{-10}$ |
| Glue | $< 2.2 \times 10^{-10}$ | $< 8.2 \times 10^{-10}$ |

The final dimensions of the $TeO_2$ crystals were within the strict tolerance of (50.00±0.05) mm per edge, with an average face flatness of under 0.01 mm. This ensured correct and reproducible thermal and mechanical coupling with the tower support structure. The square faces of the final crystals are parallel to crystallographic planes (0 0 1), (1 1 0) and (1 -1 0). The two crystal faces parallel to the crystallographic plane (0 0 1) are hard, and the remaining four are softer. More details on the crystal production can be found in [34].

The crystals were triple vacuum-packed in order to reduce any surface recontamination risks caused by radon and radon-progeny exposure (see Figure 3) [32, 33]. After production, the crystals were shipped to Italy by sea to minimize activation by cosmic rays [35] and stored underground at LNGS in vacuum sealed boxes in nitrogen flushed cabinets. Once at LNGS, a few crystals from each production batch were instrumented and tested cryogenically in order to check their radiopurity and bolometric performance [36]. These cryogenic tests, which were performed on both the CUORE-0 and CUORE crystals, demonstrated $^{232}$Th bulk and surface contamination levels of $< 2.1 \times 10^{-13}$ g/g and $< 2.0 \times 10^{-9}$ Bq/cm$^2$ (90% C.L.), respectively, and $^{238}$U bulk and surface contamination levels of $< 5.3 \times 10^{-14}$ g/g and $< 8.9 \times 10^{-9}$ Bq/cm$^2$ (90% C.L.), respectively.

### 3.2 Copper support structure

The copper used for the detector structure constitutes the largest inactive mass close to the crystals. The CUORE-0 copper support structure (which includes frames, columns and wire trays, as shown in Figure 2) has a mass of approximately 3.5 kg, while the tower shield (which surrounds the tower at the 10 mK stage and serves as thermal radiation protection) has a mass of approximately 10 kg (see Figure 13, right). Given the large amount of copper surface area facing the detector, we adopted aggressive cleaning procedures to minimize surface contamination, especially contamination by alpha emitters.

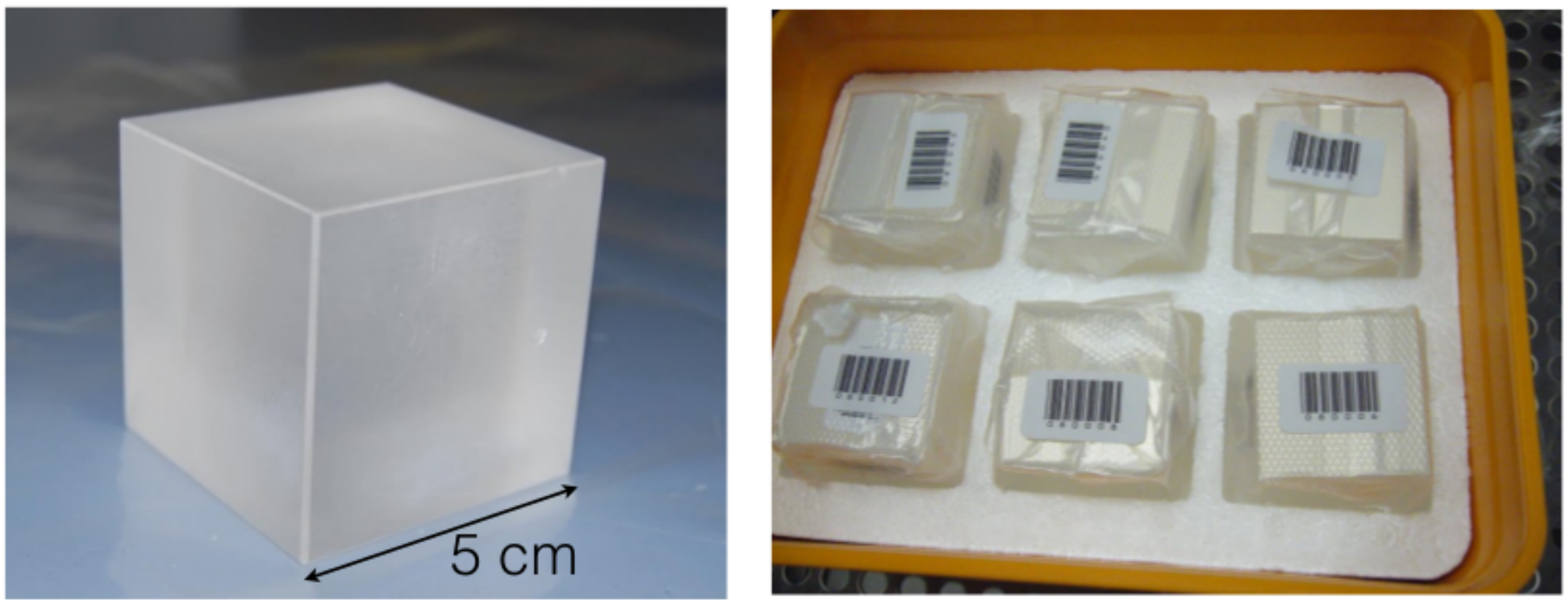

**Figure 3**. Left: A $TeO_2$ CUORE crystal. Right: Six $TeO_2$ crystals packed for transportation and storage: the crystals are triple vacuum-packed in plastic bags and they are stored in a vacuum packed box.

The detector components (i.e. frames, columns, shield) were machined from a high-purity Electrolytic Tough Pitch (ETP1) copper alloy, produced by Aurubis under the name NOSV and cast by the same company. NOSV copper was selected for its low hydrogen content and its extremely low bulk contamination levels; the upper limits are $5.0 \times 10^{-13}$ g/g for $^{232}$Th [37] and $5.3 \times 10^{-12}$ g/g for $^{238}$U (90% C.L.). To clean the surfaces of the copper detector parts, an aggressive cleaning procedure (TECM) was developed at the Legnaro National Laboratories in Legnaro, Italy, consisting of several stages: precleaning, mechanical abrasion (tumbling), electropolishing, chemical etching, magnetron-plasma etching and packaging.

During precleaning, the copper components were first manually cleaned with solvents: tetrachloroethylene, acetone and ethanol, in that order. This precleaning was designed to remove contaminants (mainly grease and oil) introduced by the machining. The components were then treated in an ultrasonic bath with alkaline soap and rinsed several times with deionized water to remove residual contamination.

The tumbling consisted of the erosion (approximately 1 $\mu$m) and smoothing out of the copper surfaces to prepare the components for the electropolishing process. The tumbling was performed in a wet environment (water and soap) with an abrasive medium of alumina powder in an epoxy cone matrix. For the thinnest components (those under 1 mm thick, including wire trays, shields and screws), the tumbling process was not performed to avoid damaging them. Instead, a soft chemical treatment was performed using a bath of ammonium persulfate. All components were then cleaned again in an ultrasonic bath with alkaline soap and rinsed in deionized water.

The electropolishing consisted of a controlled oxidation of the copper surfaces and the consequent dissolution of the generated oxide. The oxide was formed by applying a positive anode potential to the copper and was dissolved with a bath of phosphoric acid and butanol. The shape of the cathode was optimized for each type of copper component in order to make the surface erosion uniform. The electropolishing removed 100 $\mu$m of material from the copper surface, resulting in a reduction of the roughness and a mirror-like surface. As in the precleaning and tumbling steps, the components were then cleaned in an ultrasonic bath and rinsed. The chemical etching consisted of a chemically-controlled erosion of the surfaces, followed by a passivation with sulfamic acid. The components were then cleaned for the final time in an ultrasonic bath and rinsed. The final step, plasma etching, was a surface erosion produced by a plasma in vacuum. The vacuum environment prevents recontamination and promotes desorption of trace contamination remaining from the previous treatment steps. Before being sent to LNGS, each component was packed in three plastic bags under vacuum in a clean room. The plastic bags were assayed from the radioactivity point of view. Their internal contaminations were within the CUORE requirements. In order to minimize exposure to cosmic rays, which can cause copper to become activated, the raw materials were stored underground except during machining and cleaning time. In 2009-2010 we performed a bolometric test, called the Three Towers Test, to validate the copper cleaning procedure for CUORE-0 and CUORE. This test yielded upper limits of $1.3 \times 10^{-7}$ Bq/cm$^2$ (90% C.L.) for the surface contamination of the cleaned copper both for $^{238}$U and $^{232}$Th [38].

### 3.3 Germanium thermistors

Each $TeO_2$ absorber was instrumented with a Neutron-transmutation-doped (NTD) germanium thermistor. These devices convert the temperature changes in the crystals, which are typically 0.1 mK per MeV of deposited energy, to a voltage signal that is read out by the room-temperature front-end electronics. The thermistors play a crucial role in determining the bolometer signal-to-noise ratio and therefore the bolometer energy resolution and threshold.

NTD thermistors are germanium single crystals doped through exposure to thermal neutrons. Their resistance-temperature behavior roughly follows the Shklovskii-Efros law [39],

$$R(T) = R_0 \, e^{\sqrt{T_0/T}}, \tag{3.1}$$

where $T_0$ is determined by the doping level (related to the neutron fluence) and $R_0$ is determined by both the doping level and the geometry. The targets for the CUORE-0 thermistor parameters were approximately $T_0 = 4$ K and $R_0 = 1\ \Omega$, similar to the values of the Cuoricino sensors. The CUORE-0 NTD thermistors were exposed to thermal neutrons at the University of Missouri Research Reactor. Six wafers were irradiated with a nominal neutron fluence of $3.6 \times 10^{18}$ n/cm$^2$, as measured by neutron irradiation monitors. Two out of these six wafers were chosen to produce the CUORE-0 sensors. Once irradiated, the wafers were cut, optically polished, etched with nitric acid and hydrofluoric acid. The final NTDs were then boron implanted and metalized with palladium and gold at LBNL. A photograph of a CUORE-0 NTD thermistor is shown in Figure 4.

The choice of the thermistor geometry for CUORE-0 was based on experience from Cuoricino and on new constraints imposed by the CUORE-0 single-module configuration. The CUORE-0 thermistors have dimensions of $3.0 \times 2.9 \times 0.9$ mm$^3$ ($L \times W \times H$ in Figure 4). Gold pads along the sides of the thermistors also extend onto the thermistor upper face, covering two small, parallel regions of area $0.2{\times}3.0$ mm$^2$ ($P{\times}L$ in Figure 4). This wrap-around sensor design, using both lateral and frontal pads, allowed for bonding on the top of the thermistors while keeping the electrical field essentially unchanged relative to that of the Cuoricino thermistors [40], which did not have the wrap-around pads (but only the lateral ones – $H \times L$ in Figure 4). This geometry was chosen after testing numerous contact configurations compatible with frontal bonding; for a detailed discussion of the bonding procedure, see Section 4.2. The gold wire used for bonding the CUORE-0 and CUORE thermistors had a diameter of 25 $\mu$m [40]. The radioactive contamination of the gold wire was measured to be less than $1.0 \times 10^{-8}$ g/g for $^{232}$Th and less than $1.0 \times 10^{-9}$ g/g for $^{238}$U (90% C.L.).

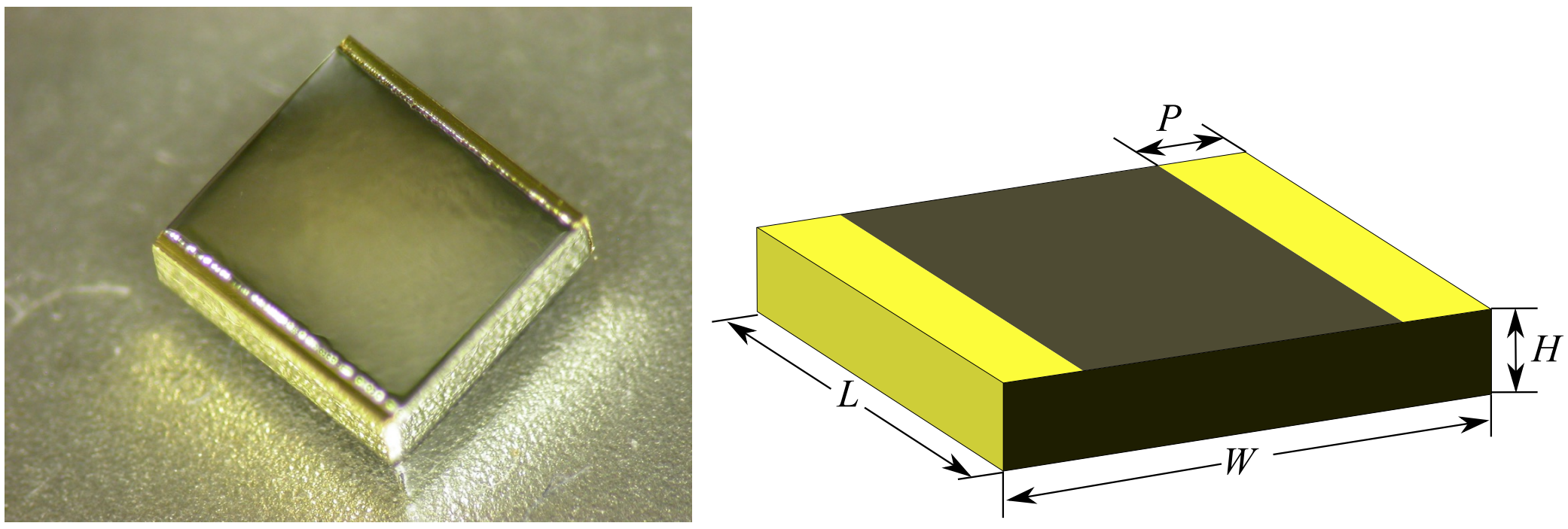

**Figure 4**. Left: Photograph of a CUORE-0 thermistor. Right: Diagram of the wrap-around thermistor geometry. Typical values for the dimensions are $L = 3.0$ mm, $W = 2.9$ mm, $H = 0.9$ mm and $P = 0.2$ mm.

After their production, five of these wrap-around NTDs were randomly selected and characterized down to 25 mK in a dilution refrigerator located in the cryogenic laboratory of the University of Insubria in Como, Italy. The resistance-vs-temperature (R-T) behaviour was compatible with that of the Cuoricino thermistors [41]. A few samples were also measured in a dedicated refrigerator located at the Centre de Sciences Nucléaire et de Spectrométrie de Masse in Orsay, France. The resistance was measured down to 12 mK, using both AC excitation and full DC voltage-current curves with consistent results. The calibration thermometers employed were routinely checked with a primary nuclear orientation thermometer. The measured values for the CUORE-0 thermistors are $T_0 = 3.84$ K and $R_0 = 1.13$ Ω, corresponding to a resistance of approximately 370 MΩ at 10 mK. The final NTD thermistors used for the detector readout showed upper limits of $1.0{\times}10^{-9}$ g/g for $^{232}$Th and $^{238}$U contaminations (90% C.L.).

### 3.4 Silicon heaters

Slow temperature drifts in thermal detectors may induce variations in their gain and spoil their otherwise excellent intrinsic energy resolution. This issue was mitigated in CUORE-0, as was done in Cuoricino, by periodically injecting a fixed-energy pulse into each crystal via a silicon heater glued to its surface. The energy pulses from the heaters emulate particle interactions and the bolometric response to these controlled events was used in offline data analysis to compensate for the effects of the temperature drifts [42].

The heaters were custom-designed $2.33 \times 2.40 \times 0.52$ mm$^3$ silicon chips on which a heavily doped meander was fabricated through the standard silicon planar process. They were manufactured by Istituto per la Ricerca Scientifica e Tecnologica (IRST, now FBK[1]) in Trento, Italy. Four aluminum pads on the resistive meander allow for a choice of resistance (100 kΩ, 200 kΩ, or 300 kΩ) depending on which pair of pads is selected (see Figure 5). For CUORE-0 300 kΩ was chosen.

[1]Fondazione Bruno Kessler

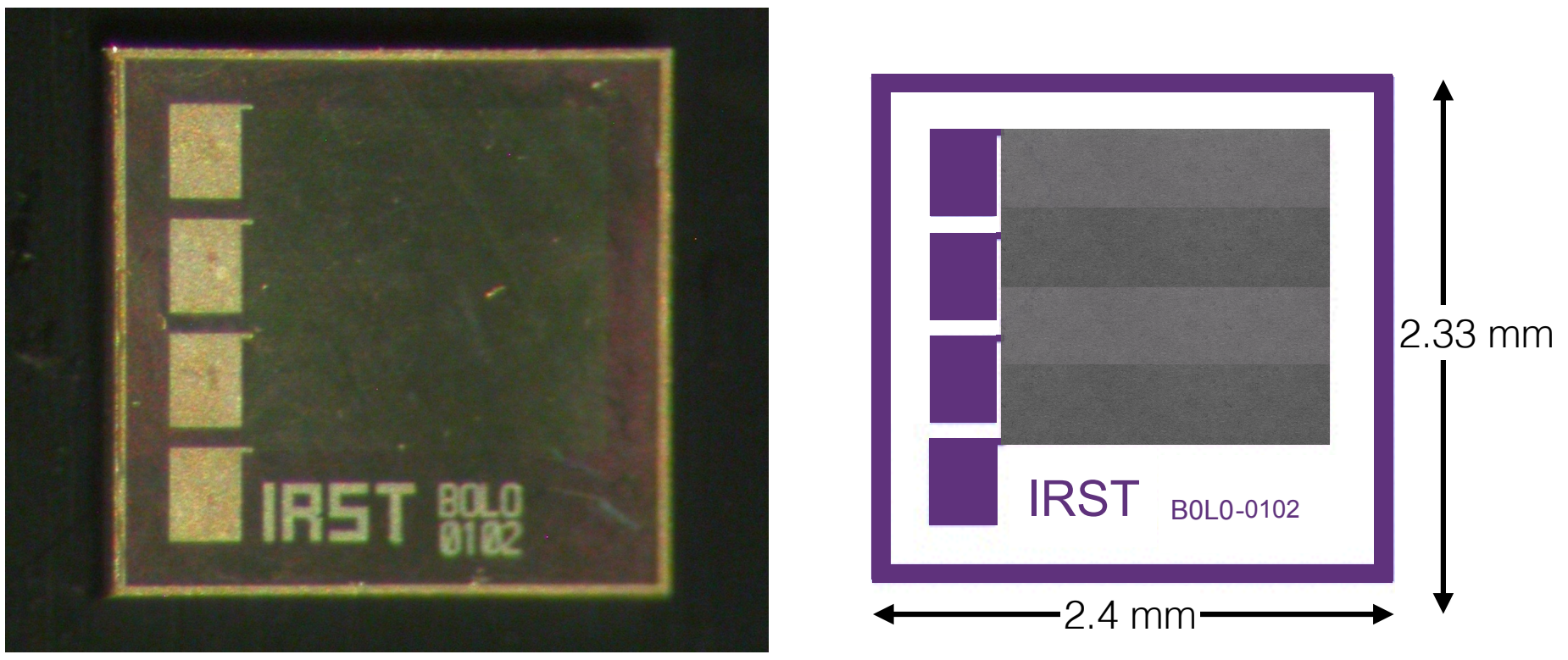

**Figure 5**. Left: Photograph of a CUORE-0 silicon heater. Right: Drawing of the heater geometry.

Each of the CUORE-0 heaters were individually characterized in a dedicated setup in the cryogenic laboratory of the University of Insubria. Initially tens of devices were characterizated down to 1.5 K, and a few samples down to 20 mK. These tests showed that nitrogen-temperature resistance measurements — performed for all the remaining devices — were sufficient to validate the heater behavior in the operating conditions of the CUORE-0 detectors [43]. The final heaters used in CUORE-0 showed upper limits of $8 \times 10^{-11}$ g/g and $1.7 \times 10^{-10}$ g/g for $^{232}$Th and $^{238}$U contamination (90% C.L.), respectively.

### 3.5 PTFE holders

The PTFE blocks play the important role of supporting the crystals inside the copper structure. Their total mass in CUORE-0 was about 290 g. The virgin PTFE was selected and validated with a long radioassay. Neutron activation analysis provided upper limits for its bulk radioactive contamination: $1.5 \times 10^{-12}$ g/g for $^{232}$Th and $1.8 \times 10^{-12}$ g/g for $^{238}$U (90% C.L.). After the production of the PTFE holders, their surfaces were thoroughly cleaned with soap and ultra-pure nitric acid, since they directly touch the crystals. After the cleaning, the blocks were vacuum sealed and stored in nitrogen until they were used in the tower assembly. The holders were designed to ensure reproducible mechanical and thermal coupling to the crystals, taking into account the differential thermal contraction of the copper, $TeO_2$ and PTFE itself. The three types of PTFE pieces used in the CUORE-0 tower are shown in Figure 6, and the position of those holders in the tower is in Figure 2.

### 3.6 Cu-PEN cables

As described in Section 3.3 and Section 3.4, each of the 52 crystals of the CUORE-0 tower was instrumented with a thermistor and a heater. The thermistor signals were read out and the heaters were powered through a set of 1.4-m long, 80-$\mu$m thick copper-insulator tapes with a polyethylene naphthalate (PEN) substrate (hereafter "Cu-PEN tapes"). A pattern of traces was etched into the 17-$\mu$m thick copper layer. The design constraints on the Cu-PEN tapes include that they must have low background, negligible cross talk, negligible microphonic pickup and that they be compactly packed during the tower assembly process. A few sample Cu-PEN strips were counted with high-purity germanium detectors and showed a contamination of less than $4.4 \times 10^{-10}$g/g for $^{232}$Th and $1.1 \times 10^{-10}$g/g for $^{238}$U (90% C.L.).

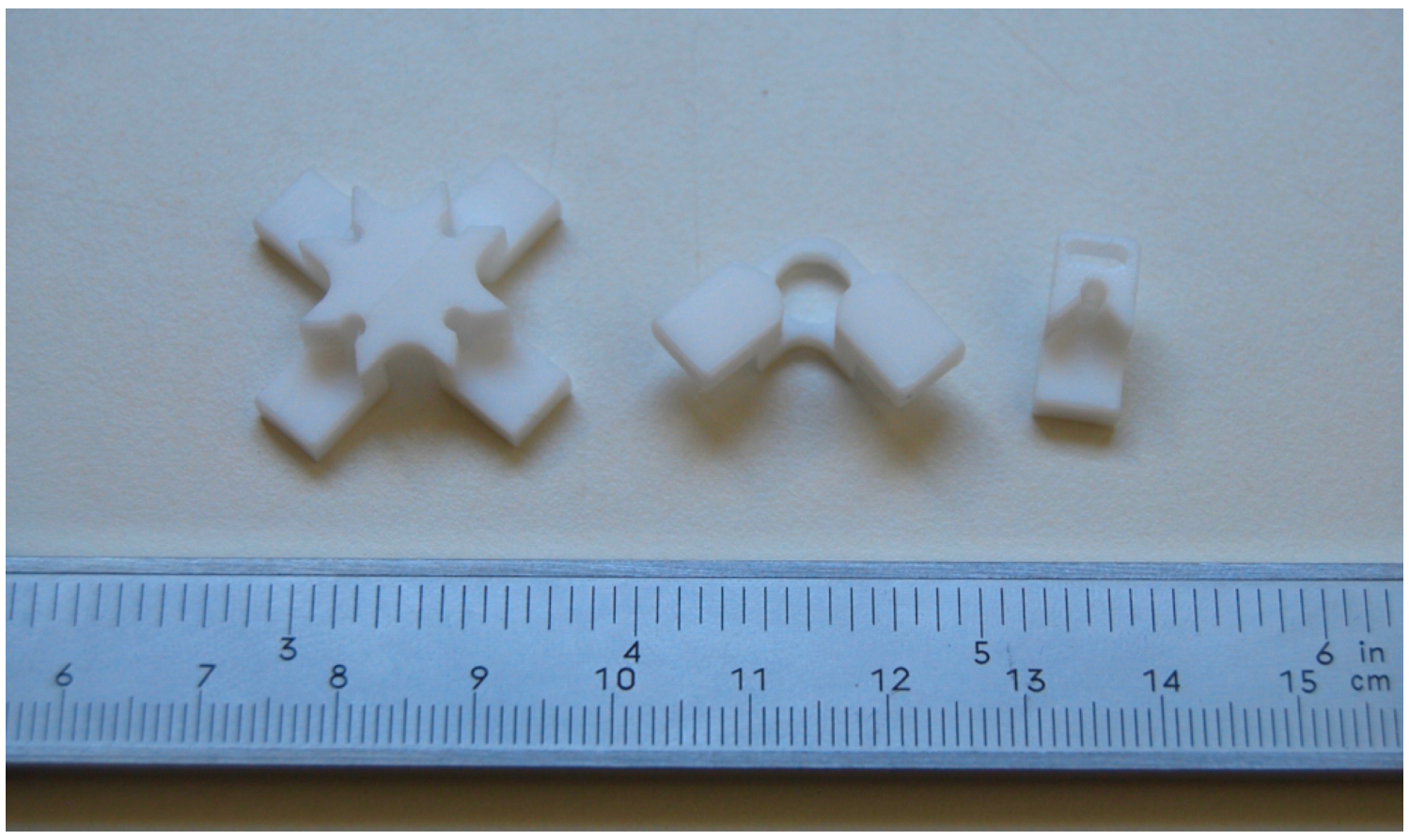

**Figure 6**. The three types of PTFE holders used for the tower construction.

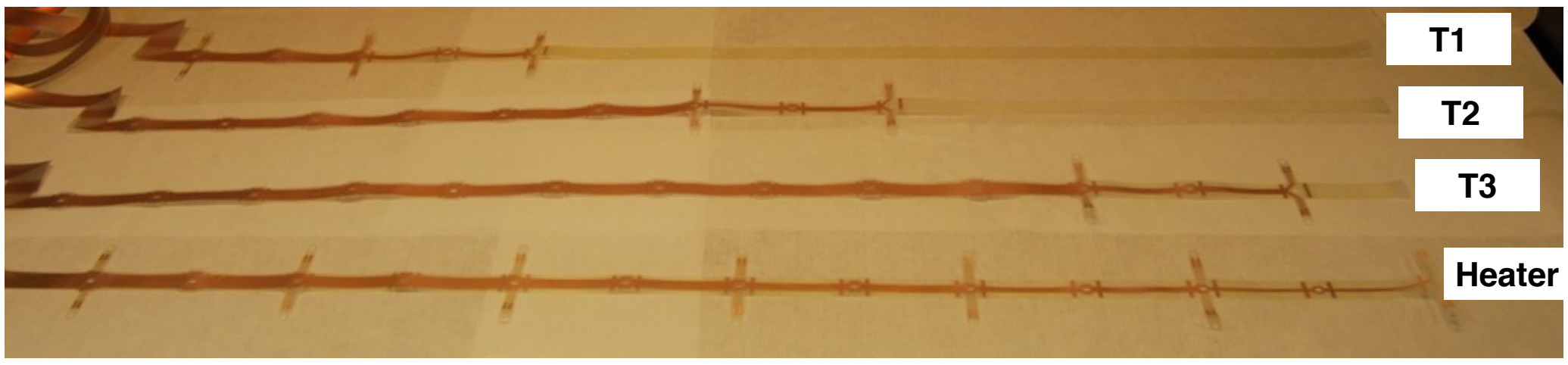

**Figure 7**. The four types of CUORE-0 Cu-PEN tapes. The picture shows the bottom ~80 cm of the tapes that were located on the sides of the tower.

The CUORE-0 tower was wired with two packs of tapes, each used to read signals from half of the tower. Each pack contained three tapes connected to the 26 thermistors on half of the tower and a fourth connected to the heaters of the same 26 crystals. These four tapes were packed together, one stacked on top of the other, with grounding tapes inserted between each for electrical shielding. The two packs were positioned and glued to copper wire trays, running on two opposite sides of the tower, such that the pads used for bonding thermistors and heaters were properly positioned at each floor of crystals in the tower. Figure 7 shows the different tape shapes: T3 for the thermistor signals from floors one to four (counting from the bottom of the tower), T2 for thermistor signals from floors five to eight, T1 for thermistor signals from floors nine to thirteen, and Heater to supply power to the heaters. The heater tapes contain two separate channels that connect each of the two columns of 13 heaters in parallel. Figure 8 shows a detailed view of the pads onto which the gold wires were bonded.

The signals from the CUORE-0 bolometers were read by differential amplifiers (detailed in Section 6). Since the common mode noise can be easily removed by these amplifiers, the Cu-PEN tapes were designed to minimize the distance between the two traces of each differential pair. This causes any mechanical vibrations to affect equally both elements of the pair. The Cu-PEN tapes were tested and characterized using the Time Domain Reflectometry (TDR) technique to check the integrity of the electrical links. The parasitic conductance in vacuum was checked using a custom-built apparatus, coupled with a commercial electrometer [44]. The differential layout pattern reduced the signal crosstalk to 0.024% and introduced a low parasitic capacitance of 26 pF/m and a parasitic impedance larger than 200 GΩ [45].

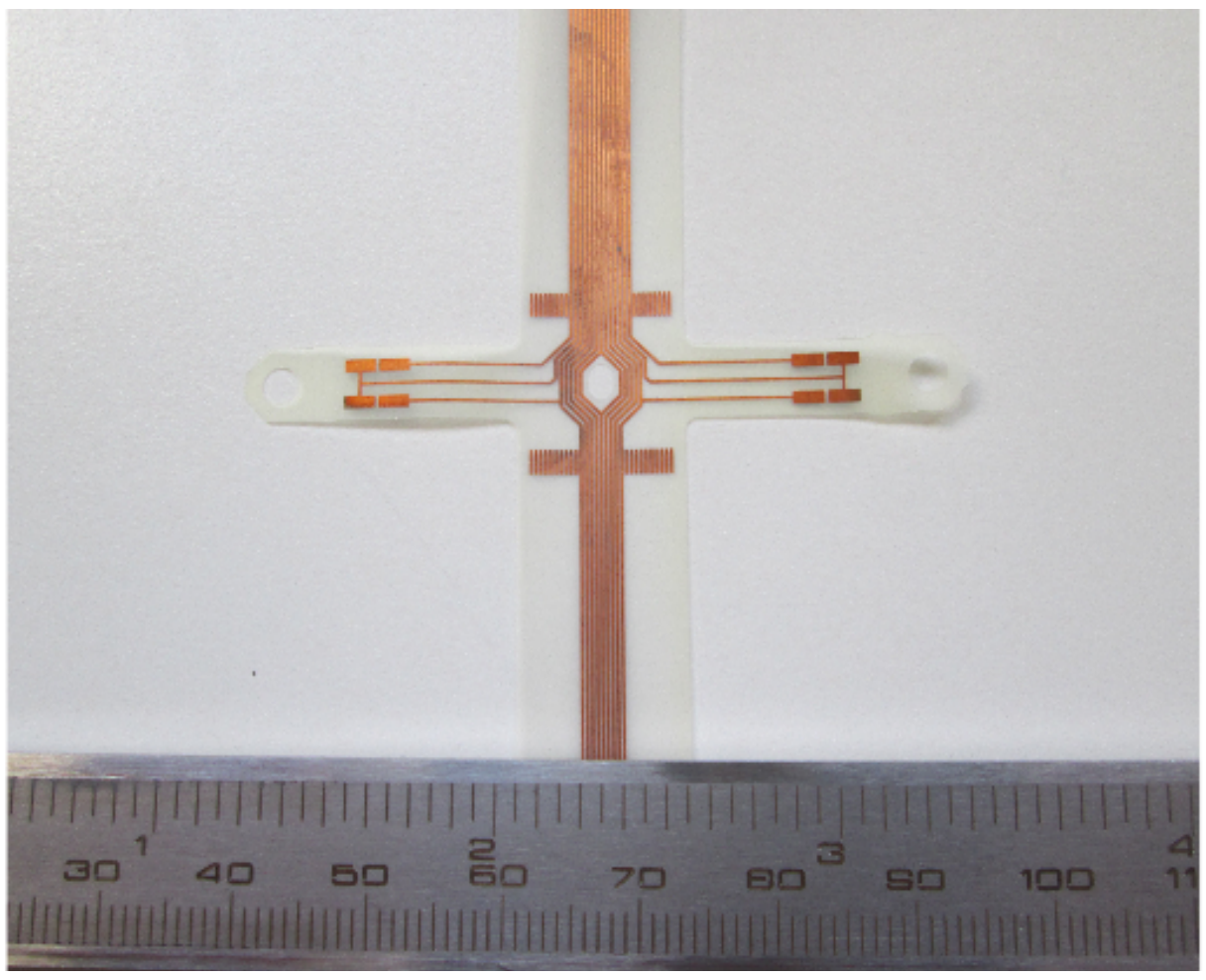

**Figure 8**. Detailed view of copper pads onto which the gold wires were bonded. The width of each copper track is 0.2 mm with a pitch of 0.4 mm. The total width of the tape is 13.4 mm, with a 1.0-mm minimum margin on each side.

## 4 Detector construction

CUORE-0 was the first detector tower built using the new techniques and assembly line developed for CUORE. All activities were carried out in a dedicated class 1000 (ISO 6) cleanroom located in the underground CUORE hut at LNGS. The cleanroom contained glovebox-enclosed systems for assembling the towers in radioclean conditions under nitrogen atmosphere. The assembly procedure used two separate workstations: one for gluing chips to crystals (see Figure 9, left), and one for building and instrumenting the towers (see Figure 9, right).

### 4.1 Gluing of semiconductor chips to crystals

The gluing operations were performed in a workstation consisting of a single glove box in order to keep the detector parts under constant nitrogen flux. Each $TeO_2$ crystal was instrumented with one thermistor and one heater using a matrix of glue dots to provide the mechanical and thermal coupling. The $TeO_2$ crystals, glue, and semiconductor chips have different thermal contraction coefficients thus the glue was distributed into dots rather than a continuous film to avoid detachment of the chips or fracturing at the chip-crystal interface due to thermal stress during the cooldown. A nine-dot matrix for thermistors and five-dot matrix for heaters were shown to provide high thermal coupling and a low risk of temperature-induced stress fractures [21]. The glue used was Araldite Rapid glue, a bicomponent epoxy produced by Huntsman Advanced Materials. This particular epoxy was selected for its rapid curing time (∼1 hour), low radioactivity (less than $2.2 \times 10^{-10}$ g/g for $^{232}$Th and $8.2 \times 10^{-10}$ g/g for $^{238}$U) and good thermal conduction. Furthermore, its cryogenic performance has been well established in previous $TeO_2$ bolometer experiments [12–15].

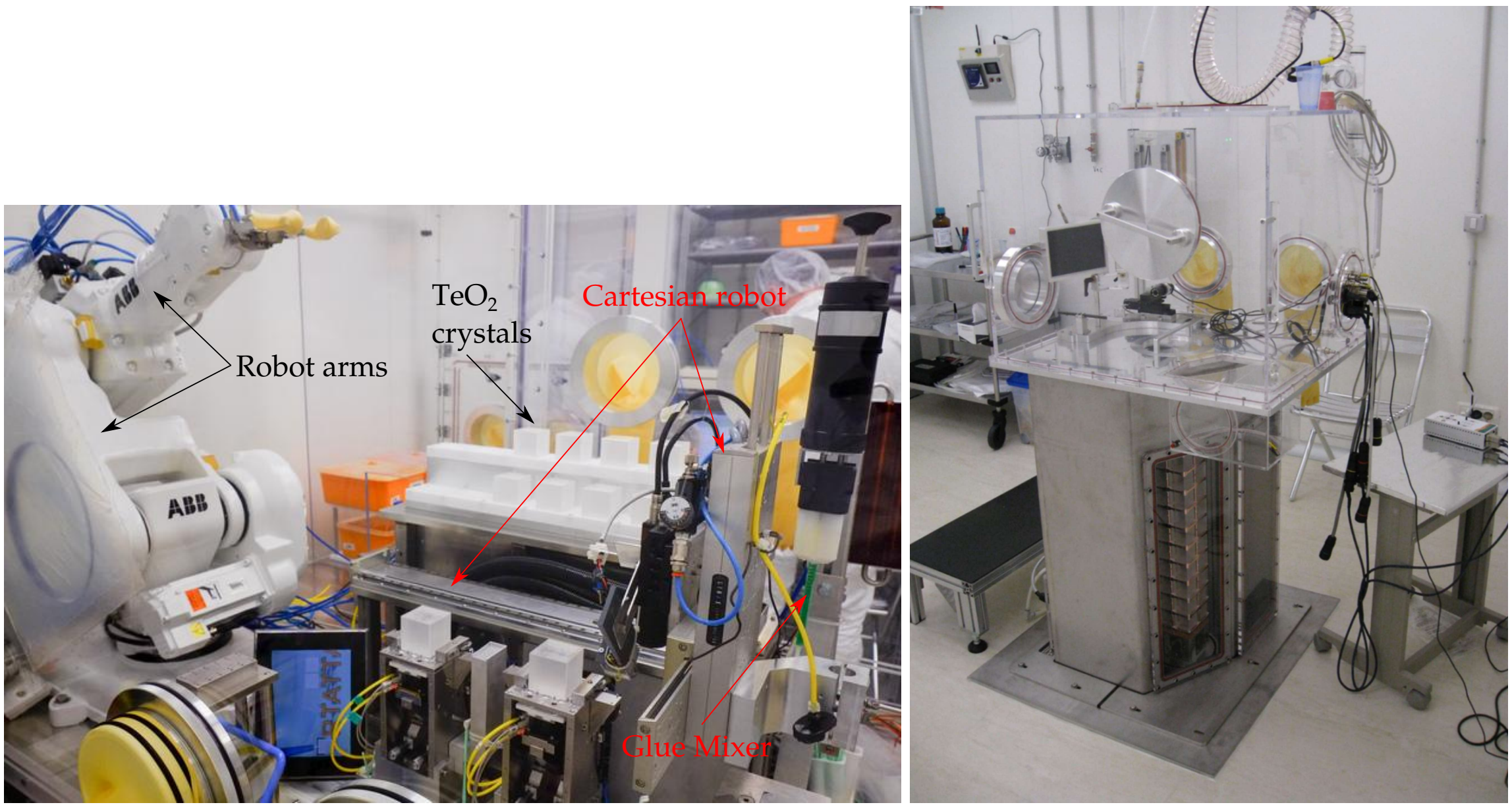

**Figure 9**. Left: The CUORE gluing workstation. Right: The CUORE assembly workstation with a glove box mounted on it.

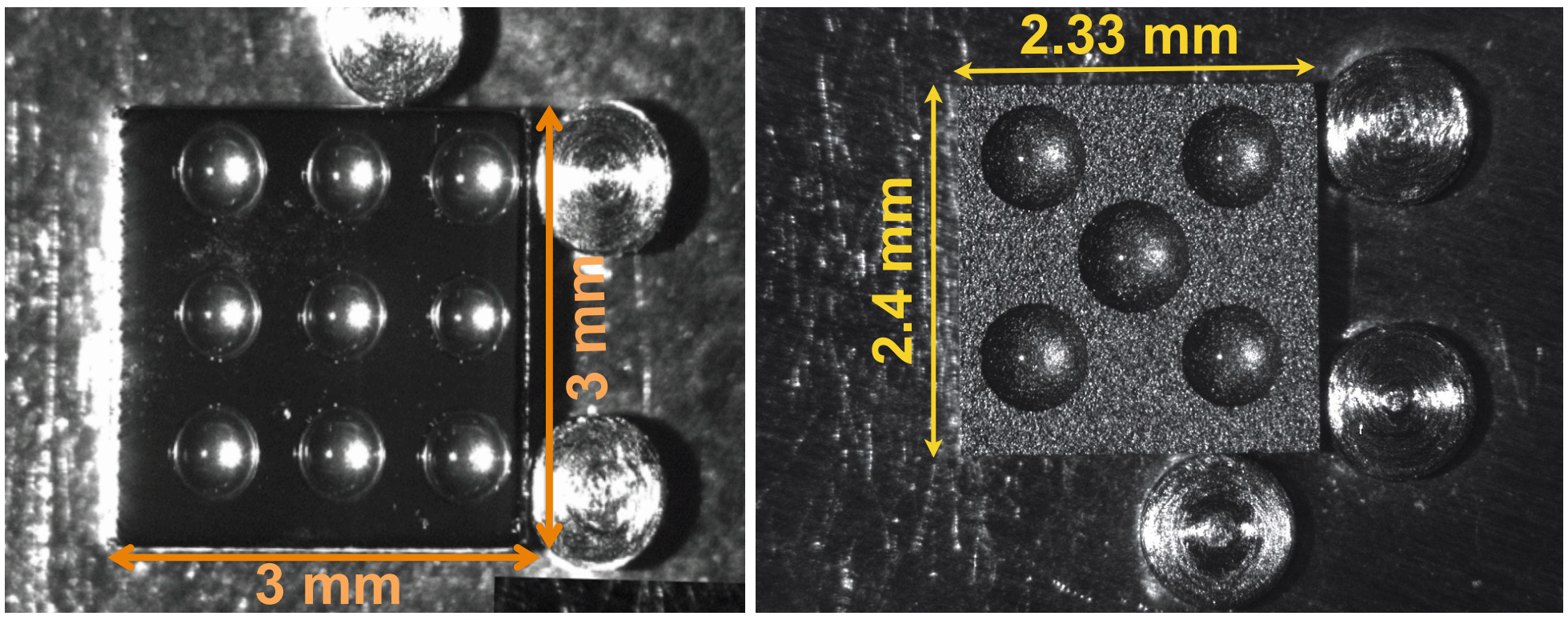

**Figure 10**. Glue-spot matrices deposited on a thermistor (left) and a heater (right) using the CUORE gluing workstation.

The quality of the chip-to-crystal glue coupling is an important factor in bolometer performance. In Cuoricino the gluing was a manual, labor-intensive procedure that caused wide variability in signal shapes among the bolometers. For CUORE-0 and CUORE, a semi-automated system was developed to achieve far more precise and reproducible chip-absorber couplings. This system included a six-axis articulated robotic arm to lift and position the crystals, and a three-axis Cartesian robot to dispense glue dots on the semiconductor chips via a pneumatic dispenser [41].

The gluing process for each crystal was as follows. First, an upturned thermistor and heater chip were placed on a positioning device. Next, the gluing sequence was initiated as epoxy components were forced from a dual-cartridge epoxy dispenser through a disposable static mixer with pneumatic pressure to achieve a consistent 1:1 mixing ratio. A small amount of the mixed epoxy was then quickly injected into a syringe attached to the Cartesian robot, which deposited the glue dots on the semiconductor chips (see Figure 10). The system photographed the glue dots and displayed the images on a nearby computer terminal for inspection. If the dots were deemed satisfactory, the operator authorized the system to proceed with attaching the chips to a crystal. The robot then retrieved a crystal from a storage shelf and placed it on a cradle above the chips, and the system lowered the cradle to bring the bottom crystal face into contact with the glue spots at a calibrated distance from the chips. The positioning device used vacuum suction to maintain the chips at a $50 \pm 5$ $\mu$m separation from the crystal face throughout the gluing process. The chips (both thermistor and heater) were glued to a hard face of the crystal. The glue dots were typically ~600 $\mu$m in diameter prior to coupling with the crystal and ~800 – 900 $\mu$m in diameter after; in the case of the thermistor, this corresponds to a total glue volume of ~0.2 $\mu$L.

To be compatible with the short pot life of the glue the entire gluing process for each sensor – from mixing to final chip attachment – was engineered to take three minutes. The timing of all steps was controlled by the system to ensure consistent, high-quality glue connections. As a precaution, no activity was performed inside the glove box for the first 30 minutes after the chips were attached to each crystal, and the glued crystal remained on the positioning device to cure for a minimum of 1 hour before being moved to a resting station. After the gluing process was complete, the crystals were placed inside vacuum-sealed container boxes, removed from the glove box, and placed inside nitrogen-flushed cabinets to await assembly into towers.

### 4.2 Tower assembly

The CUORE tower assembly line consisted of a single workstation and a suite of interchangeable, task-specific glove boxes and equipment mounted atop it in sequence (see Figure 9, right) [31]. Below the station's working plane was an airtight stainless-steel enclosure continuously flushed with nitrogen that could be opened directly into the glove box. This enclosure served as a garage for safely storing a tower during glove box exchanges or pauses in assembly activity. After each glove box was installed atop the workstation, it was flushed for several volume exchanges before the garage cover was opened.

The first step in assembling the CUORE-0 tower was to physically construct it from copper, PTFE, and 52 crystals with semiconductor chips attached. For this operation, a special glove box was designed to facilitate the clean ingress of tower components and their assembly by hand. The tower was built one floor at a time and gradually lowered into the garage as it grew in height in order to maintain a constant operation level. Some of the steps in the assembly of the tower are shown in Figure 11.

Once the tower was built, a different glove box was installed on the workstation to enable attachment of the two packs of Cu-PEN tapes (described in Section 3.6) on opposite sides of the tower. The cables were first glued to the wire trays using Araldite Standard bicomponent epoxy, which provided a rigid copper backing, and after curing overnight the two cable assemblies were attached to the tower. During the assembly operations, ~60 cm of Cu-PEN tapes exceeding the tower's height were kept rolled up inside PTFE cylinders (shown on the top of the tower in Figure 13).

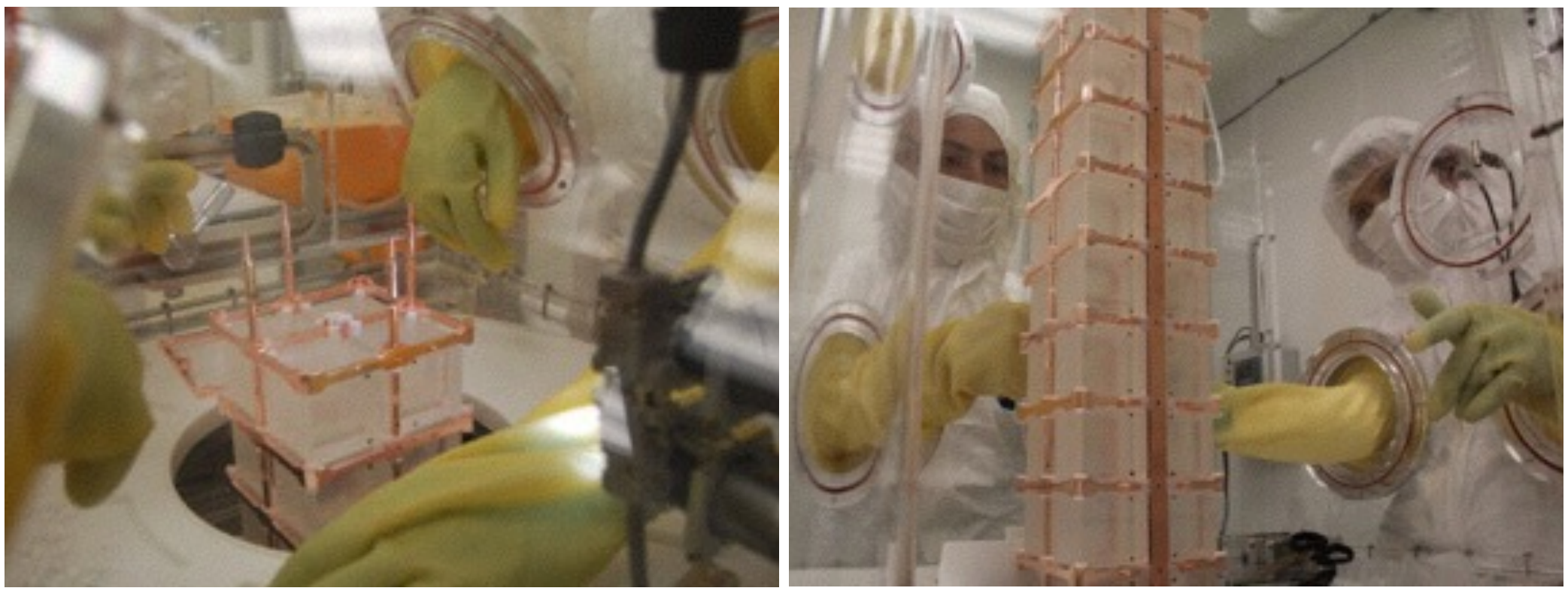

**Figure 11**. Left: the CUORE-0 tower during its construction. Right: the CUORE-0 tower during the installation of the read out cables.

Next, a specially modified Westbond 7700E manual wire bonder, which was mounted on motor-driven rails, was used to make the electrical connections between the chips and the readout traces. Each 25-$\mu$m-diameter gold-wire connection was first ball-bonded to a chip pad and then wedge-bonded to a copper-trace pad, and the wedge bond was reinforced with a security ball bond (see Figure 12). Two wires were bonded for each electrical connection to provide redundancy. The bonding process was well-defined and repeatable so that the lengths of each wire were as uniform as possible.

One of the 52 NTD thermistors in the CUORE-0 tower could not be bonded due to imperfect curing of the epoxy beneath the corresponding copper trace, which rendered the surface too soft to achieve a bond. One of the 52 heaters (on a different crystal) could not be bonded for the same reason. Failing to bond a thermistor meant the bolometer could not be readout, while the bonding failure of the heater meant that the thermistor could still be readout but heater-based stabilization of the bolometer gain was not possible.

After the wire bonding of the CUORE-0 tower was complete, protective copper covers were installed over the Cu-PEN tapes and the tower was enclosed in a cylindrical copper radiation shield (see Figure 13, right). The radiation shield consisted of a segmented cylinder fully surrounding the tower, coupled to the bottom and top plates. In CUORE there will be a single radiation shield for the entire array and not one for each single tower.

## 5 Experimental setup and shielding

CUORE-0 was located in Hall A at LNGS at a depth of ~3600 m.w.e. The muon and neutron fluxes in Hall A are $3 \times 10^{-8}$ $\mu$/(s·cm$^2$) [46] and $4 \times 10^{-6}$ n/(s·cm$^2$) [47]. The CUORE-0 detector was assembled underground in the cleanroom of the CUORE hut, close to the CUORE-0 hut. The fully assembled tower was moved from the CUORE cleanroom to another cleanroom on the first floor of the CUORE-0 hut. Here the tower was attached to the same dilution refrigerator that was used to run the Cuoricino detector.

During the transportation from the CUORE cleanroom to the CUORE-0 cleanroom, the tower was enclosed in a plastic cylindrical canister (see Figure 13, right) filled with nitrogen gas. Nitrogen flux was maintained for as long as possible while connecting the detector to the cryostat to minimize exposure of the tower to air.

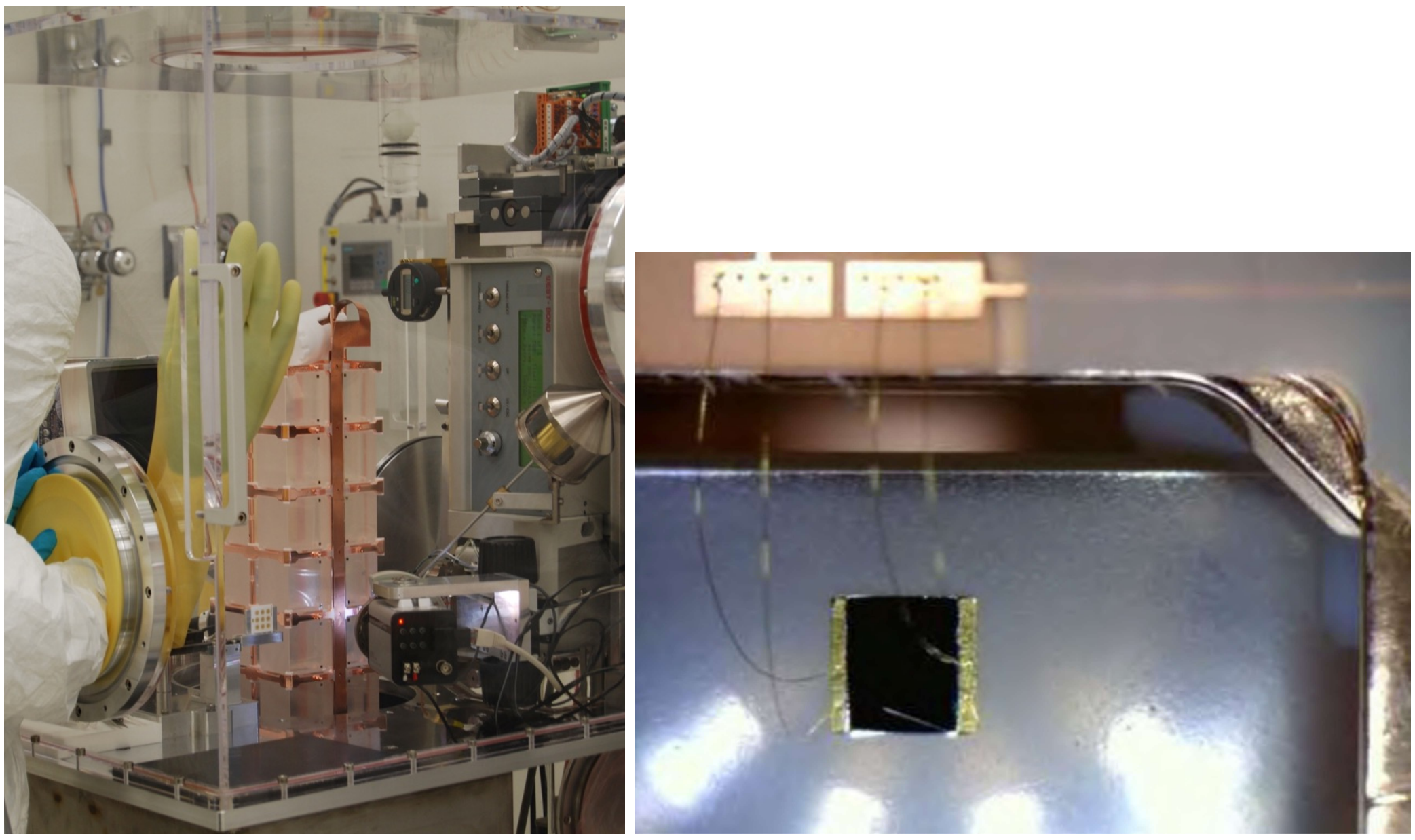

**Figure 12**. Left: The assembly workstation during the bonding operations. The bonding machine is on the right of the picture. Right: Detail of the bonding connection of a NTD thermometer. The gold pads of the thermistor are connected to the copper pads of the Cu-PEN cable through gold wires.

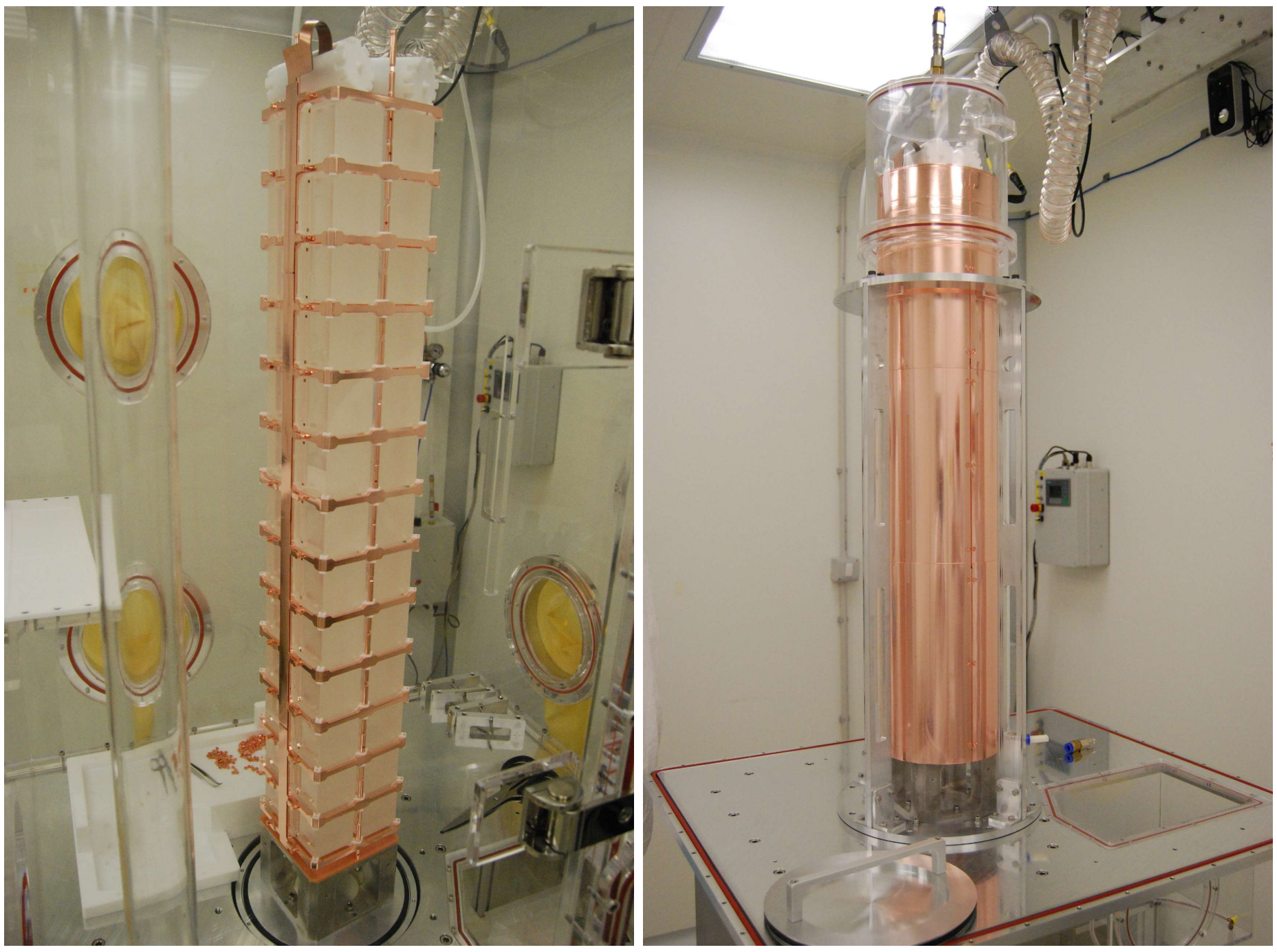

**Figure 13**. Left: The completed CUORE-0 tower. Right: The shield-enclosed CUORE-0 tower inside its nitrogen-flushed storage and transport canister.

In order to isolate the detectors from vibrations caused by the cryogenic equipment, the tower was connected to the cryostat with a stainless-steel spring fixed to a plate thermalized at 50 mK [48]. The tower was thermally linked to the mixing chamber of the dilution refrigerator by two $80 \times 10 \times 0.05$ mm$^3$ copper foils, connected to the bottom of the mixing chamber and the top copper plate shield of the tower. Diagnostic thermometers and heaters were also installed on the bottom and top of the tower. The Cu-PEN tapes were unrolled from the PTFE cylinders and plugged into zero insertion force (ZIF) connectors mounted on Cu-Kapton[2] boards on the top of the mixing chamber (see Figure 14). The Cu-Kapton boards connected the Cu-PEN tapes to 2-m-long links to the top of the cryostat. The links were made of twisted niobium-titanium (NbTi) wires interwoven with NOMEX[3], thermalized at different temperature stages of the cryostat. At the top of the cryostat, the wires were soldered to multiple 27-pin connectors produced by Fischer Connectors. Outside the cryostat vacuum, twisted cables, about 1 m in length, carried the signals from the top of the cryostat to the front-end electronics. The residual capacitance and parasitic impedance of the NbTi-NOMEX ribbon cables were measured using a commercial LCR meter and commercial electrometer [49]. The results showed a capacitance of about 100 pF/m and a negligible level of cross-talk.

The tower was enclosed inside the set of nested cylindrical coaxial vessels which make up the thermal radiation shields of the cryostat. In order of installation, they are identified as the 50 mK, the 600 mK, and the 4 K shields. The 4 K shield also formed the walls of the Inner Vacuum Chamber (IVC). The apparatus was then lowered through a trapdoor in the floor of the cleanroom into a vessel holding the liquid helium main bath. The main bath vessel formed the inner walls of the Outer Vacuum Chamber (OVC) and was located inside a larger vessel that formed the outer walls of the OVC. Thus, the OVC comprised the space between the main bath external vessel (at 4.2 K) and the OVC external vessel (at room temperature), as shown in Figure 15. The OVC contained 5 more coaxial cylindrical vessels along with several hundred thin aluminized mylar sheets, which acted as super insulation to reduce the radiation from the external vessel at room temperature. All of the vessels in the cryostat were made of oxygen-free high-conductivity (OFHC) copper.

The cryostat was supported by, and hung through, the center of a square aluminum plate, which was supported by three pneumatic level mounts affixed to the top of the external lead shield (see Figure 15). The level mounts enabled easy adjustment of the inclination of the support plate in order to hang the detector nearly vertically inside the cryostat and prevent the tower's innermost shield from touching the thermal shield at 50 mK. The height of the support plate above each level mount was monitored by a 1-$\mu$m-resolution Heidenhain ST 3088 length gauge affixed to the external shield. The level mounts were lowered or raised by emptying or filling them with nitrogen gas via computer-actuated valves. During data taking, the cryostat was generally "reverticalized" every 2–3 weeks to compensate for the gradual deflation of the level mounts under load.

The CUORE-0 detector used the same shielding that was previously used for the Cuoricino experiment. The tower was surrounded by a 1.4-cm-thick Roman lead cylindrical shield, whose $^{210}$Pb activity was measured to be below 4 mBq/kg [50]. A Roman lead disc, 10 cm thick and 17 cm in diameter was placed just above the tower to shield the detector from the intrinsic radioactivity of the dilution refrigerator; similarly a disc 8 cm think and 19 cm in diameter was below the tower to shield from activity in the laboratory floor. The cryostat was also shielded by two layers of lead externally. The first external layer was 10 cm of modern lead with a measured $^{210}$Pb activity of 16 ± 4 Bq/kg, while the second layer was 10 cm of higher-activity modern lead measured at 150 ± 20 Bq/kg. The external lead shielding was surrounded by an acrylic glass anti-radon box flushed with nitrogen gas to reduce Rn contamination. Finally, a 10-cm-thick layer of borated polyethylene was used as neutron shielding. The entire apparatus was enclosed in a Faraday cage to minimize local electromagnetic interference. The walls of the cage were then covered with sound and vibration absorbing material to reduce microphonic noise. A full drawing of the CUORE-0 shields is shown in Figure 15.

[2]Kapton is a registered trademark of E.I. Du Pont de Nemours and Company.
[3]NOMEX is a registered trademark of E.I. Du Pont de Nemours and Company.

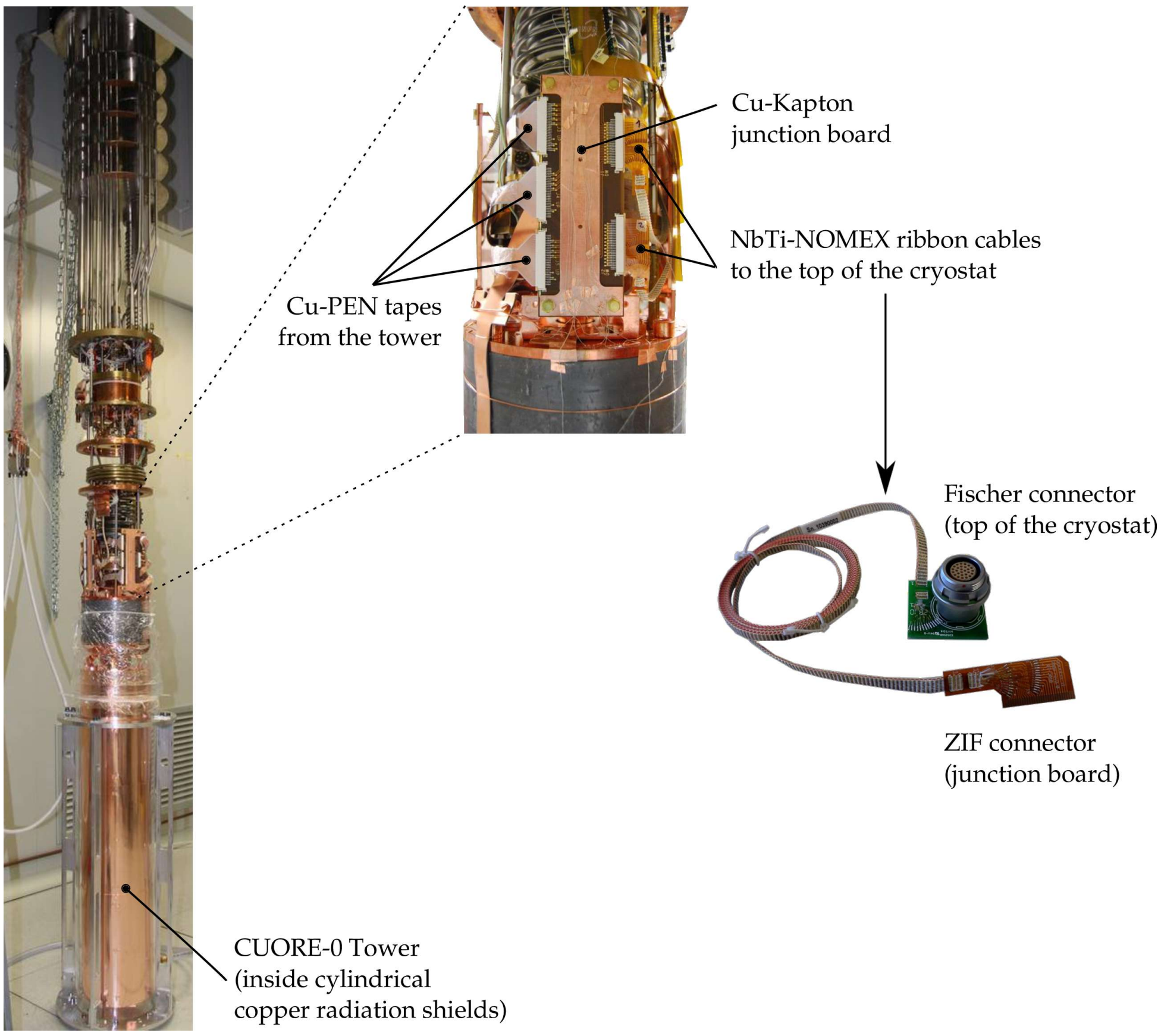

**Figure 14**. Photograph of the CUORE-0 tower attached to the dilution refrigerator, before the installation of the thermal radiation shields. Right: A zoom of a Cu-Kapton junction board, which connects the Cu-PEN cable to the NbTi-NOMEX cables which run up to the top of the cryostat.

## 6 Signal readout

The CUORE-0 electronics provided a low-noise system to read out the detector signals and to monitor and optimize its performance. It consisted of six parts: front-end, antialiasing filter, pulser, linear power supply, AC/DC pre-regulator and data acquisition system. A block diagram of the system is shown in Figure 16. Many parts of the system were remotely controlled using an $I^2C$ serial communication bus [51] routed with optical fibers.

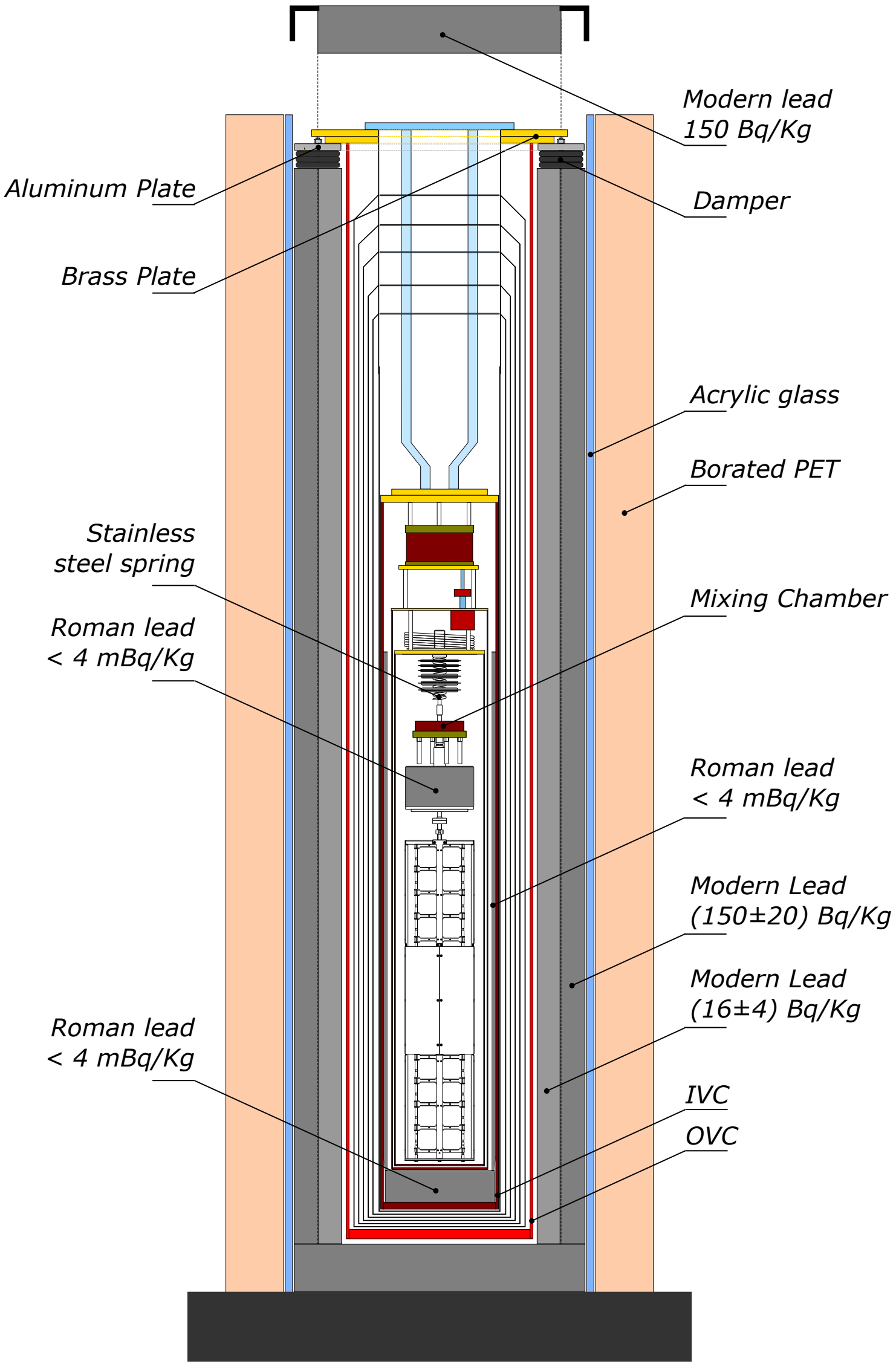

**Figure 15**. Sketch of the CUORE-0 cryostat and shielding (not to scale). The indicated activities refer to decays of $^{210}$Pb.

### 6.1 Front-end main board

When a particle interacts with the absorber, the absorber heats up and the resistance of the thermally-coupled NTD thermistor changes accordingly. Given their high impedance the NTD thermistor were biased by a DC current that transformed the resistance change to a change in voltage in the range $\Delta V_{\text{bol}}$ = (50 – 300) $\mu$V/MeV. The signals passed through an amplification system to exploit the dynamic range of the DAQ (+/-10.5V). The amplification was provided by custom Front-End

Boards (FEBs). Each board consisted of two analog channels and their associated digital logic circuitry. Each channel was composed of a pair of load resistors and their differential biasing systems, a differential voltage-sensitive preamplifier (DVP) [52], and a programmable gain amplifier (PGA) [53]. The biasing system and the PGA were located on the FEB, while the load resistor and the DVPs were accommodated on a separate daughter board. Each FEB was equipped with a Digital Control Board, a daughter board that was able to configure all the parameters that influenced the analog signal processing chain and communicate with the remote controller. Each Digital Control Board accommodated three complex programmable logic devices (CPLDs): one for each amplification channel and one for remote communication. The FEBs were physically located above the cryostat inside the Faraday cage, very close to the cryostat output connectors.

The bias current supplied by the bias generator was sourced across the thermistor through a pair of low-noise large-value thick-film load resistors on a dedicated daughter board (LVR board). The board accommodated a pair of 27 GΩ resistors ($R_{\text{load}}$ = 54 GΩ) and, by using a latching relay, these could be connected in parallel with two 6.8 GΩ resistors, lowering the total load resistance to 10.9 GΩ. This feature was used when large biasing currents were needed in the bolometers (e.g. for thermistors with low impedance). Since the parallel noise from the load resistors was one of the main contributions to the total noise, resistors that exhibited low $1/f$ noise when biased at the current levels required by the detectors were selected [54].

Each channel's bias was individually adjusted, so that each bolometer could be operated at its optimal working point, as described in Section 7.3. As such, each channel was equipped with a dual programmable attenuator that allowed for an adjustable bias voltage across the load resistors. This programmable bias generator was also able to reverse the polarity of the bias voltage to cancel any offset. The nominal range of the bias voltage was from -10 V to +10 V. To extend the dynamic range to 60 V (from -30 V to +30 V) the system also accepted an external voltage input.

The DVP was designed to satisfy stringent requirements in terms of noise and stability. It contained a pair of low-noise silicon JFETs at the input. The closed loop gain was 220 V/V. The differential voltage-sensitive configuration allowed the inputs to float, which made the detector biasing compatible with DC coupling and suppressed common-mode disturbances and cross-talk. The DVP was accommodated on a daughter board and had a series noise of less than 7 nV/$\sqrt{\text{Hz}}$ (5 nV/$\sqrt{\text{Hz}}$ average) at 1 Hz and 3 nV/$\sqrt{\text{Hz}}$ at high frequencies (white noise) [52]. Preamplifier parallel noise was less than 0.1 fA/$\sqrt{\text{Hz}}$ thanks to the small gate current of the semi-custom JFET (a few tens of fA) [55]. This value was 6 times less than that of the LVR board, making the preamplifier contribution negligible.

The DVP was equipped with an offset corrector designed to compensate the thermistor bias voltage, within the range −80 to +80 mV at the input, in order to maintain the signal at the end of the amplification within the range −10.5 to +10.5 V. The correction current was supplied by a circuit based on a programmable 12-bit multiplying digital-to-analog converter.

An important requirement for the preamplifier was to have a very small temperature drift. The intrinsic drift was mainly due to the gate-to-source voltage mismatch of the two input JFETs. To compensate for the intrinsic drift, a static current was injected into a suitable preamplifier node. This current, proportional to the absolute temperature, was derived from a forward-biased diode, after an attenuating network calibrated for each preamplifier. This correcting method reduced the drift at the output to a fraction of $\mu$V/°C.

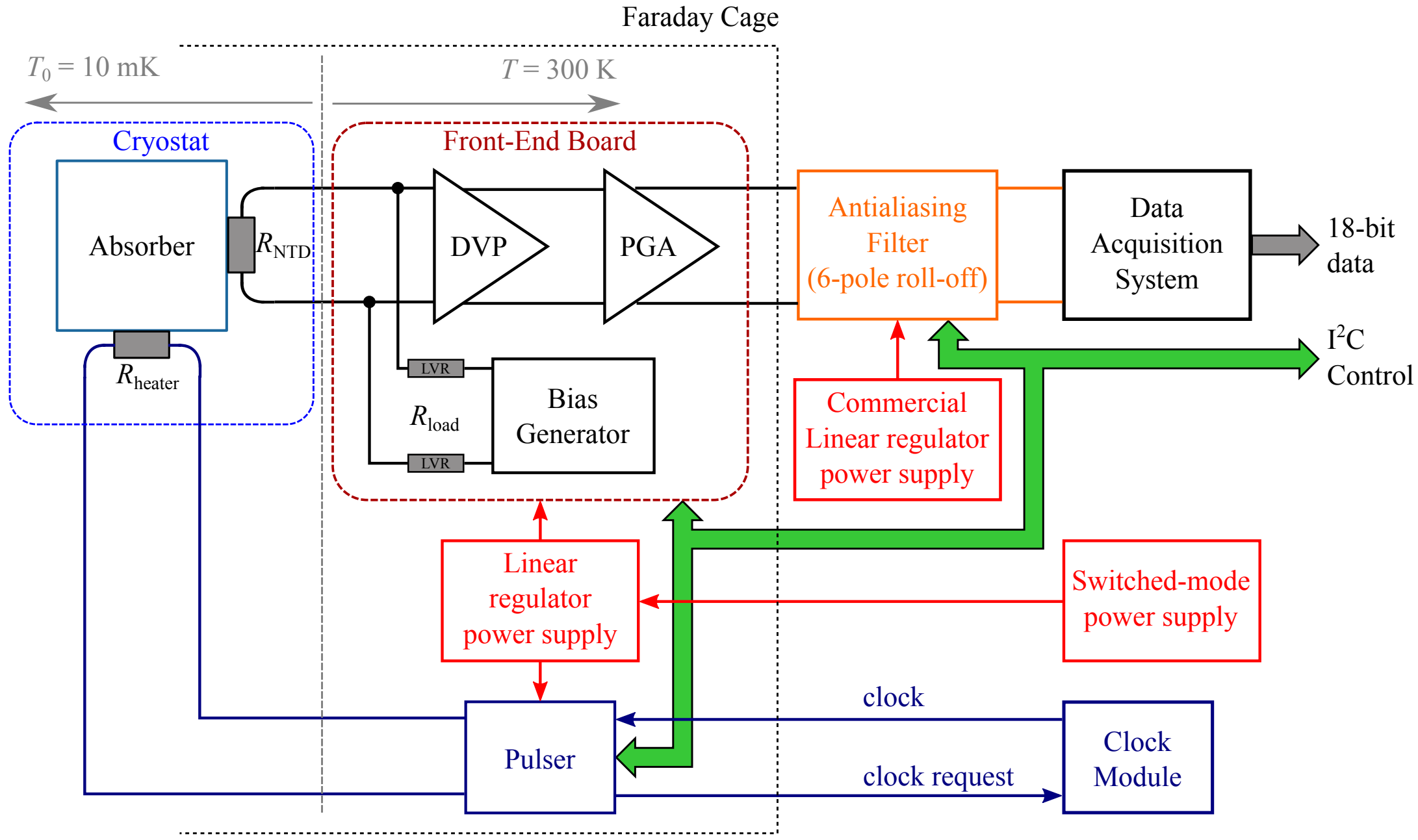

**Figure 16**. CUORE-0 electronics block diagram.

The outputs of the preamplifiers fed the differential inputs of the PGA present on the FEB. Its gain was set with 5 bits of resolution, giving the FEBs an end-to-end gain of 220–5000 V/V.

### 6.2 Thomson-Bessel antialiasing filter

The outputs of the FEBs were connected to an antialiasing filter before entering the data acquisition system. The purpose of this stage was to reduce the aliasing noise in the out-of-band frequency region and the sampling artifacts. The filter, an active six-pole Thomson-Bessel Low-Pass Filter (LPF) with a rolloff of 120 dB/decade, was located in a dedicated Faraday cage close to the data acquisition system far from the FEBs, minimizing any possible pickup after the filtering. A Thomson-Bessel filter was chosen because it preserves the shape of the input signal even when its rolloff is close to the signal frequency bandwidth [56]. Each board consisted of three channels, each of which had four programmable frequency bandwidths (from DC to 8, 12, 16 and 20 Hz) to accommodate the signal characteristics of each detector. A CPLD was used to set the cut-off frequency, enable and disable the filter, and communicate with the remote control system.

### 6.3 Temperature stabilization system

Intrinsic instabilities in the complex cryogenic setup can spoil the detector energy resolution if they cause the temperature of the detector tower to fluctuate. To compensate for these instabilities, a proportional-integral-derivative (PID) controller was implemented to maintain the tower temperature as close as possible to the the desired setpoint during the extended duration measurements which were typical during stable operations. The PID controller used in CUORE-0 was based on the system developed for Cuoricino [57]. The PID loop used an amplifying chain composed of a preamplifier and programmable gain amplifier to implement the proportional term. The input to the PID loop was the temperature measured on an NTD thermistor (i.e. the voltage across a stabilization thermistor) while the control output was generated by a transconductance amplifier $g_m$. This amplifier implemented the integral and derivative terms, and its output current fed the stabilization heater.

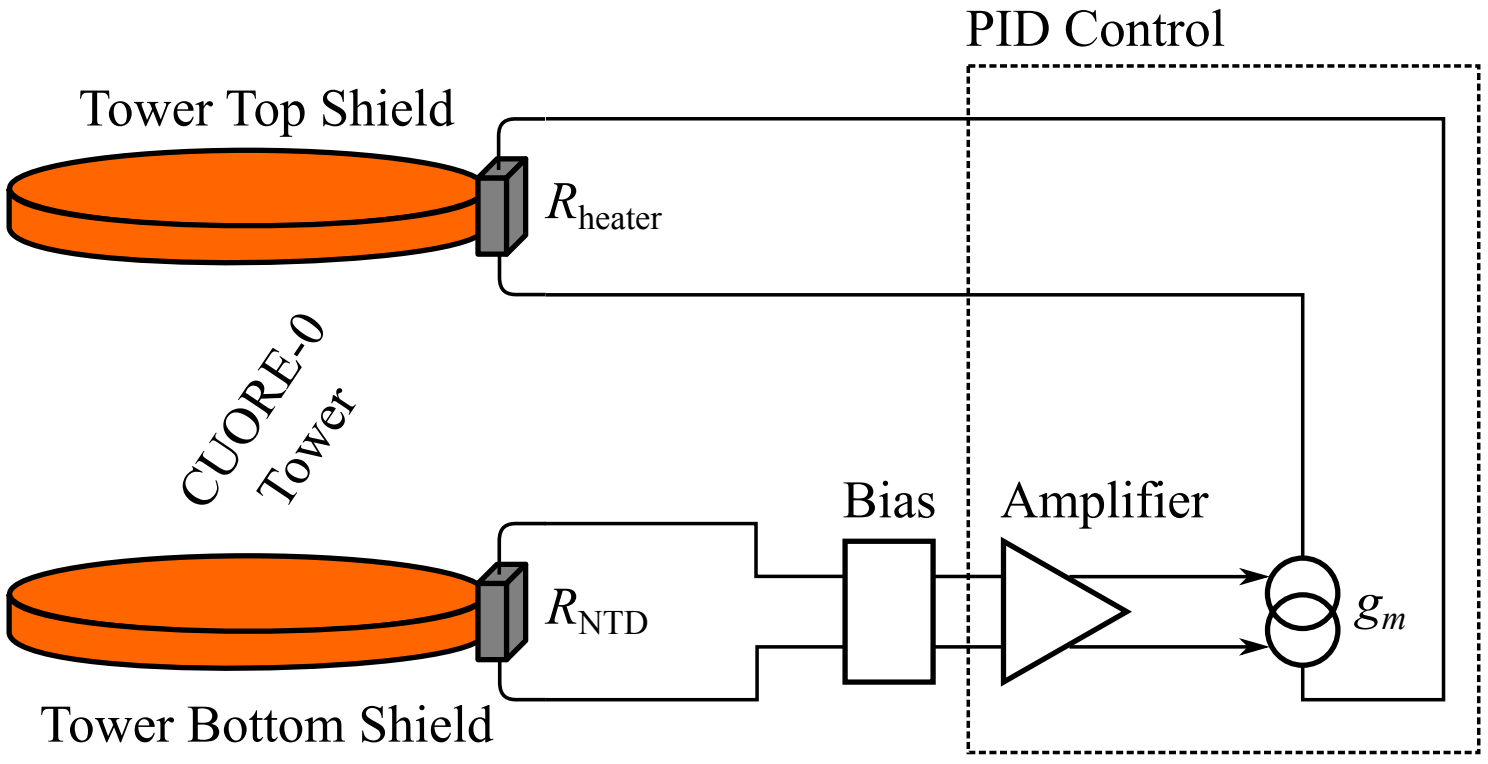

**Figure 17**. Temperature stabilization set-up of the CUORE-0 tower.

In CUORE-0 there were two stabilization channels. The first was used to stabilize the mixing chamber, and the related thermistor and heater were placed on the mixing chamber itself. The second was used to stabilize the tower, and the thermistor was placed on the bottom copper shield while the heater was placed on the top copper shield (Figure 17). Both thermistors and heaters used were of the same type as those used for the crystals.

The temperature was stabilized by the offset-correcting circuit of the remotely programmable front-end channel, which autogenerated the reference operating point for the feedback loop. Using this configuration, it was possible to minimized the effects on the bolometer detectors due to the instability of the cryogenic apparatus. Overall a stability of the detector baselines of better than 0.5 keV/day was achieved. More details can be found in [57].

### 6.4 Response stabilization pulser system

As described in Section 3.4, in order to perform thermal gain stabilization, a fixed amount of energy was periodically delivered to the detector by silicon heater resistors glued to each crystal absorber, inducing responses similar to that of signal events [42, 58]. This technique allowed us to monitor the pulse amplitude variations and correlate these variations with the detector baseline drifts in the data analysis.

The heater pulses were generated by a custom four-channel pulser board [59], whose output was designed to be stable to within 1 ppm/°C. The heaters for each column of the tower were connected in parallel, so each channel controlled 13 heaters. The board generated square pulses whose amplitude and duration were independently configurable for each output channel by a CPLD-based, remotely-controlled digital circuit. The duration of the pulse was configurable between 0.1 and 25.5 ms with a step size of 0.1 ms. For CUORE-0, pulses 1 ms wide were generally used; this width is much shorter than the typical rise time of the signal pulses generated by particle interactions.

A precise 10 MHz clock module was used to determine the pulse duration. This clock was placed outside the Faraday cage to avoid ground loops and minimize electromagnetic interference. The pulser boards communicated with the module through two optical links: "clock" and "clock request". The clock request line was used to enable the precision clock only during the pulse generation.

### 6.5 Power supply system

The stable operation of bolometers strongly depends on the stability of their power supply, which must exhibit very low noise and very low drift. The CUORE-0 power supply system was composed of a custom power distribution chain for the devices that required very stable behavior (e.g. FEBs and the response stabilization system) and a commercial linear regulator power supply (Agilent 6627A DC power supply) for the devices that were less sensitive to the stability of the supply (e.g. the antialiasing filter boards).

The custom supply chain was composed of a Vicor switched-mode power supply filtered by a Vicor VI-RAM[4] passive low pass filter and an active filtering module, whose purpose was to attenuate the switching noise to negligible levels. The switched-mode power supply outputs drove several custom linear regulator power supplies, one for the pulser crate and one for each FEB crate. Each custom board was a linear bipolar (±10 V) voltage supply/reference featuring low noise, high power supply rejection ratio (PSRR) and high stability of approximately 1 ppm/°C [60]. This board also provided to every FEB the voltage reference reference used to generate the adjustable bias voltage of the thermistors. Each board included safety circuits to protect itself, the electronic equipment it supplied, and the detector in general.

### 6.6 Data acquisition system

The CUORE-0 data acquisition (DAQ) system digitized the bolometer signals coming from the antialiasing Bessel filters and stored the data on disk for offline analysis. It also provided auxiliary features, including tools for online monitoring and software interfaces to the front-end electronics, Bessel filters, and pulser board. The DAQ hardware consisted of five National Instruments (NI) PXI-6284 digitizer boards in a NI PXI-1036 chassis. Each board digitized 16 differential channels with 18-bit vertical resolution over the symmetric range −10.5 to +10.5 V and was operated at a sampling frequency of 125 samples/s. The chassis was connected to the DAQ computer by an optical link provided by a NI PXI-PCI8336 controller. The bolometer waveforms were sampled continuously and the complete stream was transferred to the computer for processing and storage.

The data were stored in two forms, a continuous waveform format that stored 100% of the data, and a triggered format which selected a subset of the waveforms for further analysis. The following triggers were defined for the standard analyses:

1. *Signal trigger.* A signal event was identified when the slope of the waveform exceeded a given threshold for a certain amount of time. Because the pulse shape and the signal to noise ratio could change from bolometer to bolometer, the trigger parameters were configured independently for each channel.

[4]VI-RAM is a trademark of Vicor Corp.

2. *Noise trigger.* This trigger fired periodically (every 200 s), simultaneously on all channels, regardless of the presence of pulses. It was used to understand the noise behavior of the detector in the absence of signals.

3. *Pulser trigger.* An event of this kind was generated when the pulser board (see Section 6.4) fired, injecting a predetermined amount of energy in the silicon heaters attached to the bolometers. When the pulser board fired, a digital signal was sent to a dedicated line on the digitizer boards, and the events from the pulsed bolometers were flagged in the DAQ system.

The triggered data were stored in ROOT [61] files. Each event contained some header information, including time since the beginning of the run, channel, event number, trigger type (signal, noise or pulser), and the waveforms of the triggered bolometer and the bolometers geometrically close to it. Nearby bolometers were defined as those from the same floor, the floor below and the floor above the triggered bolometer. For each trigger on a bolometer on the top or bottom floor of the tower, this resulted in 8 saved waveforms; for triggers on all other bolometers, 12 waveforms were saved. The recorded waveforms had a length of 5.008 s (or 626 samples) and included 1 s (125 samples) of data before the trigger. This pre-trigger waveform was used to evaluate the baseline level at the time of the trigger. ROOT files of the triggered data also included header information that was not specific to each event, such as run number, run start date, and run type (physics data or calibration). This general information was also stored in a database that could be accessed by the analysis software. Before the events were saved to ROOT files, they underwent a fast online analysis that evaluated some basic parameters such as the channel trigger rate, amplitude spectrum, baseline RMS and slope, rise time, and decay time. These online parameters were accessible through a web-based interface which was updated every few minutes. The continuous streams of the bolometer waveforms were saved in compressed ASCII files. These files were not used in the standard data analysis, but they were used to run more sophisticated trigger algorithms [62] and for additional studies of the detector.

In addition to the core functionalities of data digitization, run control and monitoring, the DAQ system also provided software interfaces for the control of the FEB, Bessel filter and pulser electronics. The low-level communication interfaces were wrapped in a client-server infrastructure that made them easily accessible for a variety of applications, including graphical user interfaces for the display and control of the configuration parameters of the electronics devices. This infrastructure was also used for automated detector characterization procedures in which the data readout and the electronics control subsystems were used together to measure each detector's performance as a function of certain configuration parameters. Two of these automated characterization procedures are described in Section 7.3 and Section 7.4.

## 7 Detector operation and performance

The first data-taking campaign of CUORE-0, corresponding to a $^{130}$Te exposure of 2.0 kg·y, started in March 2013 and lasted until September 2013 [63]. Following this the cryostat was shut down for maintenance. Detector operations resumed in November 2013 and benefitted from reduced microphonic noise and improved duty cycle resulting from the maintenance operations. CUORE-0 col-lected data for $0\nu\beta\beta$ decay analysis until March 2015, for a total $TeO_2$ exposure of 35.2 kg·y, corresponding to a 9.8 kg·y $^{130}$Te exposure. Figure 18 shows the exposure accumulated by CUORE-0 over time.

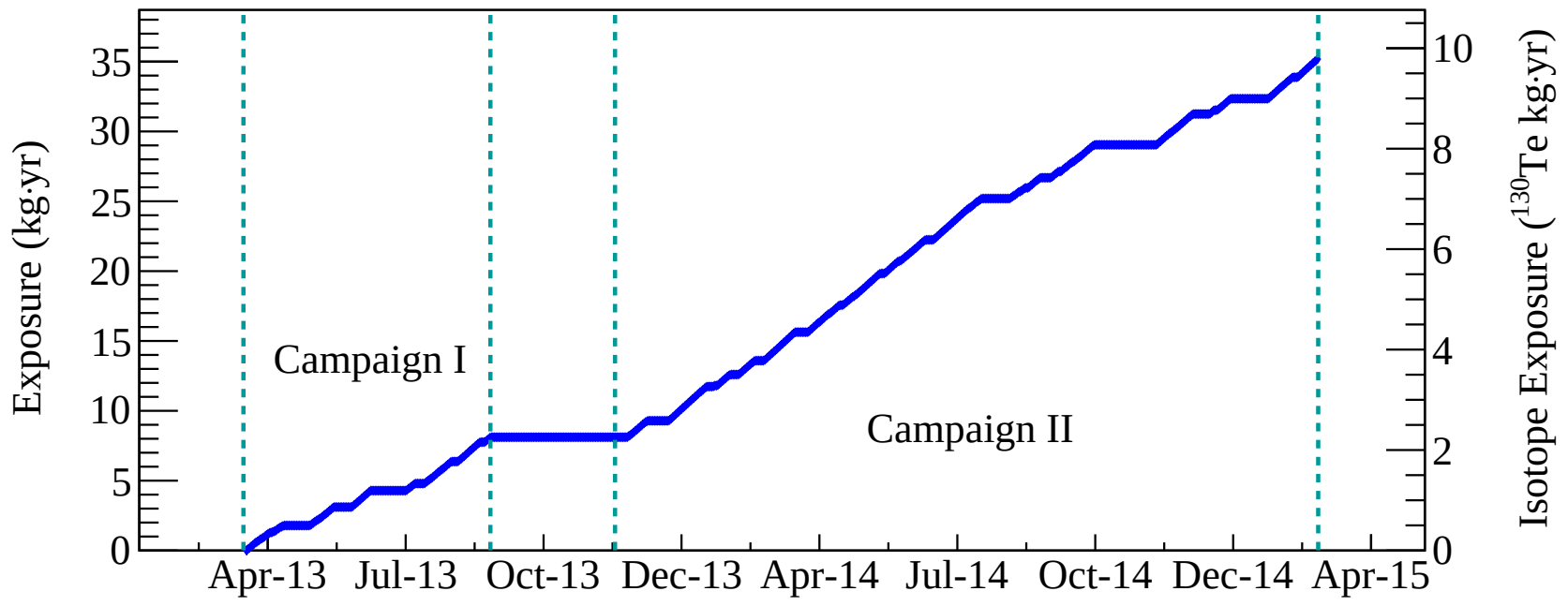

**Figure 18**. Plot of the accumulated statistics over time. The left vertical axis refers to total $TeO_2$ exposure, while the right one refers to $^{130}$Te exposure.

During the initial cooldown, the electrical connection to the silicon heater on one channel was lost. This was the only failure of the readout wiring, despite several thermal cycles during the commissioning phase. Including one heater and one thermistor which could not be bonded during the assembly of the tower (see Section 4.2), overall the detector had 49 bolometers with both thermistor and heater working properly. Using a new thermal gain stabilization technique [42], it was possible to include the two bolometers with thermistors but without heaters in the global $0\nu\beta\beta$ decay analysis, bringing the number of active bolometers in CUORE-0 to 51.

In this section we discuss the operation and performance of the CUORE-0 detector. The corresponding results on the search for $0\nu\beta\beta$ decay can be found in a dedicated paper [5], and further details of the analysis are covered in [42].

### 7.1 Data collection

The CUORE-0 data collection was organized in runs, with each run lasting approximately one day. Once every other day, data taking was paused to refill the cryostat's liquid helium bath. Each refill introduced a down-time of 2 – 3 hours. Runs were grouped into datasets, each containing about three weeks of physics data (the data collected for the $0\nu\beta\beta$ decay study). When possible, a dataset started and ended with a calibration measurement, each of which lasted approximately three days. Some datasets have only an initial calibration due to problems in the cryogenic system that interrupted the data taking before the planned final calibration.

In calibration and physics data runs, the pulser board (see Section 6.4) fired every 300 s on the bolometers. A complete cycle of ten pulses (3000 s period) was repeated for the duration of the measurements. Eight of these pulses had the same amplitude, corresponding to energies in the 3 – 4 MeV range, depending on the bolometer; these pulses were used for the offline correction of the detector gain instabilities induced by temperature drifts of the apparatus. Of the remaining two pulses of each cycle, one was at a higher amplitude, corresponding to energies in the 6.5 – 8.5 MeV range, and the other was at a lower amplitude, corresponding to the 1.5 – 2.3 MeV range. These two pulses were used to check the effectiveness of the thermal gain stabilization at other energies. In physics runs the signal trigger rate per bolometer was around 1 mHz, and it was around 60 mHz in calibration runs.

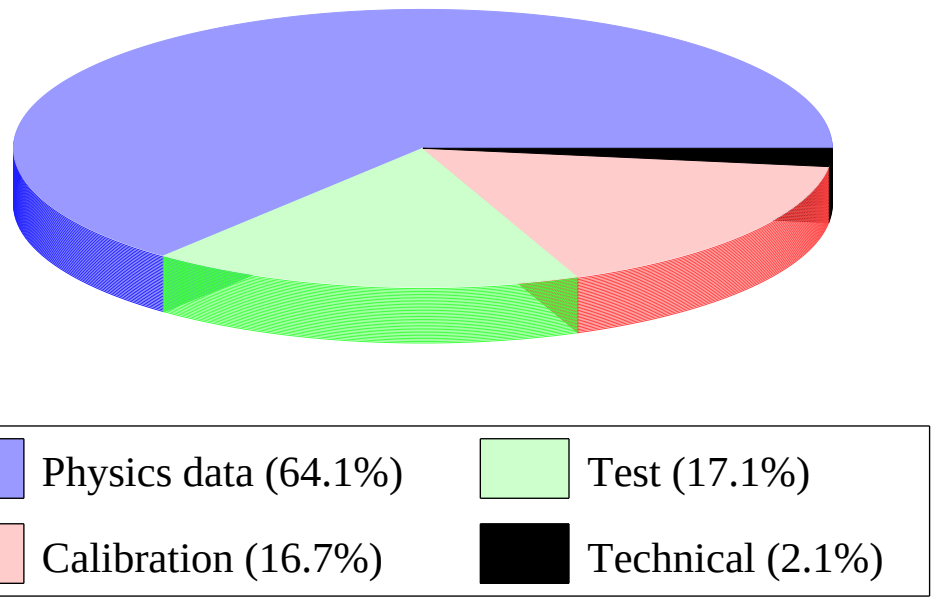

**Figure 19**. Fractions of CUORE-0 live time spent in different measurement types.

The duty cycle of the CUORE-0 detector was 78.6%, including the downtime between the two data-taking campaigns. The partition of the live time is illustrated in Figure 19, and includes physics data (64.1%), calibration (16.7%), test runs (17.1%) and technical runs (2.1%). Only physics data and calibration runs are used for data analysis. Test runs include monitoring of the detector during the usual liquid helium bath refills and measurements to optimize the cryogenic parameters of the apparatus. The remaining 2.1% corresponds to technical runs that were performed periodically. They include the optimum working point searches described in Section 7.3 and the daily working point measurements and scans with the pulser at low energy described in Section 7.4.

### 7.2 Thermistor uniformity

The CUORE-0 data allows us to verify the improvement in the uniformity of the bolometric performance achieved with the new CUORE-style assembly procedure compared to the previous experiment, Cuoricino. It is well known that the coupling between the sensors (thermistors and heaters) and the crystal absorber is critical for optimizing the bolometric behavior. In order to better understand the degree of uniformity of the detector response, the distribution of the CUORE-0 base temperatures measured by the thermistors was evaluated and compared to that of Cuoricino. Before biasing, all of the bolometers should be at the same temperature, and thus a narrow distribution of measured temperatures is expected from reproducible uniform assembly. Non-uniformity in the thermistor-absorber coupling would broaden the temperature distribution of the thermistors. The temperature of the thermistor from Equation 3.1 can be written as

$$T(R_{\text{base}}) = T_0 \left[ \ln \left( \frac{R_{\text{base}}}{R_0} \right) \right]^{1/2}, \tag{7.1}$$

where $R_{\text{base}}$ is the resistance of the thermistor measured by the voltage drop across it in the limit of no bias current. To estimate the base temperature of each bolometer, the $R_{\text{base}}$ was measured immediately after the detector cooldown. Identical measurements were performed during Cuoricino. Using this data and the measured values of $R_0 = 1.12\ \Omega$ for Cuoricino and $R_0 = 1.13\ \Omega$ for CUORE-0 the distribution of $T/T_{\text{avg}}$ was evaluated for Cuoricino and CUORE-0. Here $T_{\text{avg}}$ denotes the average of $T(R_{\mathrm{base}})$ over all bolometers measured in each detector. The ratio of the temperature and the average was adopted as it can be found more accurately than the absolute temperature because it depends only on $R_0$ , and only logarithmically. The result is shown in Figure 20. The RMS of the distribution is 9% in Cuoricino and 2% in CUORE-0. The narrower distribution of CUORE-0 temperatures compared to that of Cuoricino is a clear demonstration of the improvement achieved with the new CUORE-style assembly procedure described in Section 4. For reference we also report the average temperature values, evaluated using the measured values of $T_0$ = 3.84 K for CUORE-0 and $T_0$ = 3.41 K for Cuoricino. Using these values, we obtain an average temperature of 10.05 mK for CUORE-0 and of 8.02 mK for Cuoricino.

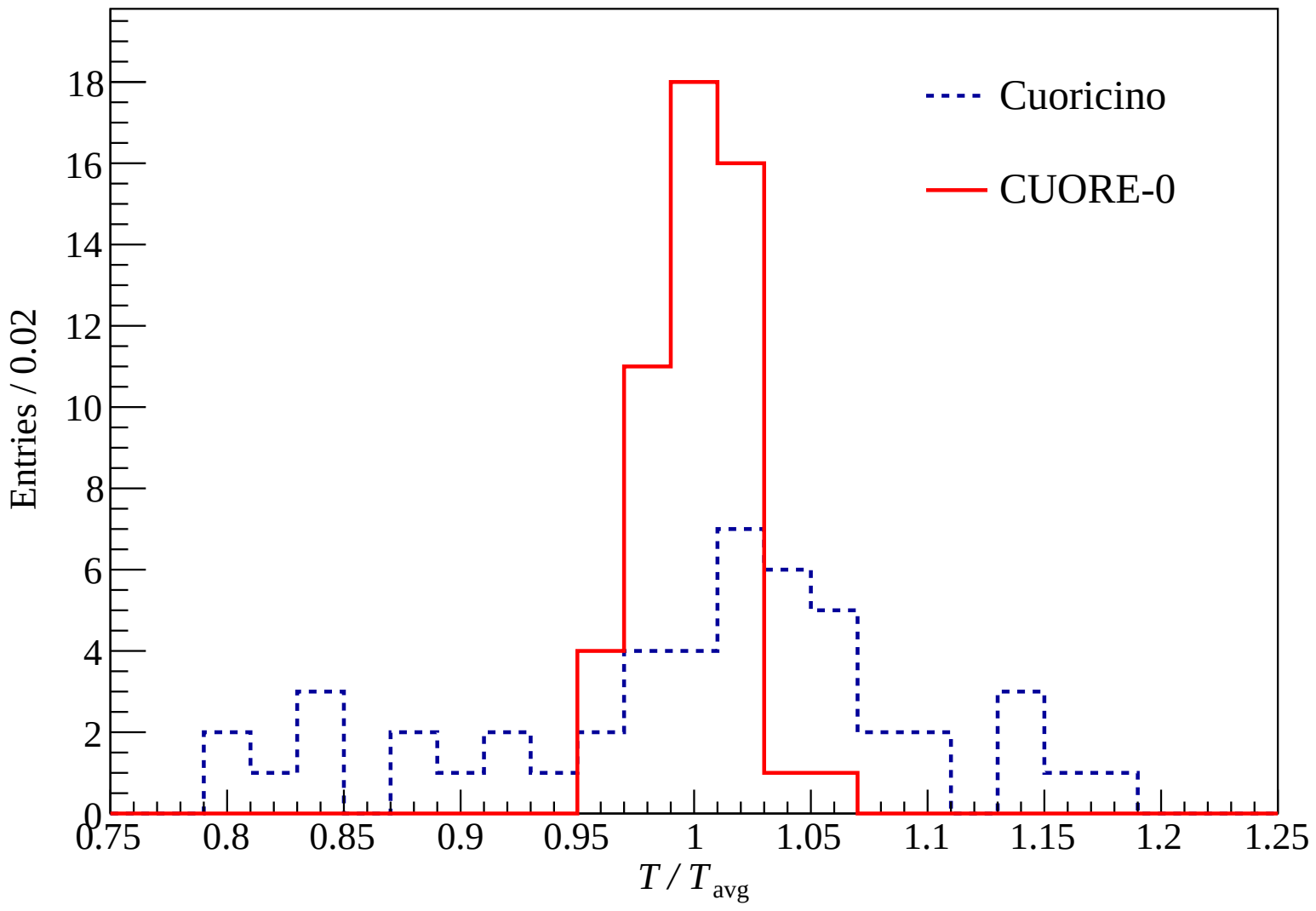

**Figure 20**. Comparison of the base temperatures of the bolometers normalized to their average temperature for CUORE-0 (red line) and Cuoricino (blue dashed line).

### 7.3 Bolometer working points

For a given working configuration of the cryogenic apparatus, the response of a bolometer depends on the value of the bias current used for its operation. For this reason, the optimal working point (i.e., the bias current that maximized the signal-to-noise ratio) was evaluated for each of the CUORE-0 bolometers at the beginning of each campaign and every time the parameters of the cryogenic apparatus changed significantly. The measurement consisted of evaluating the bolometer voltage, the baseline RMS and the amplitude of a fixed-energy reference pulse (see Section 6.4) for each value of the bias current for each bolometer. The bolometer voltage was evaluated from two baseline noise measurements with the same bias current but opposite polarity. To eliminate possible voltage offsets at the input of the readout chain, the bolometer voltage was taken to be the difference of these two output baseline voltage values divided by the known gain of the readout chain. The baseline noise RMS was evaluated as the integral of the average noise power spectrum [42]. The amplitude of a reference pulse was obtained by averaging the waveforms of several pulser events.

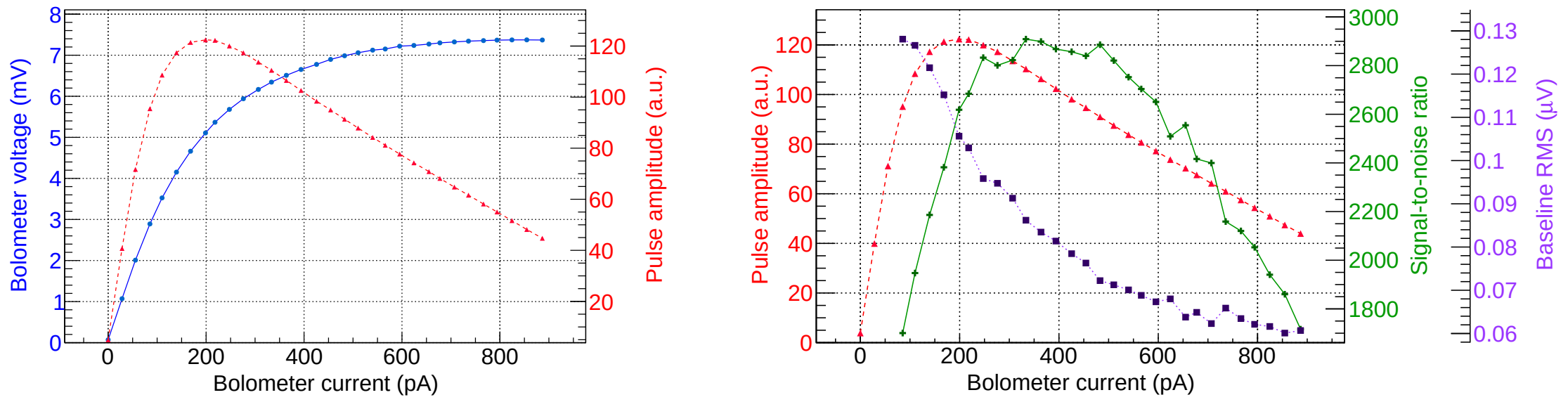

**Figure 21**. Detector characterization curves. Left: I-V curve (blue solid) and pulse amplitude curve (red dashed). Right: The same pulse amplitude curve (red dashed) with the baseline RMS curve (purple dotted) and their ratio (green solid).

Figure 21 shows the results of the optimal working point measurement for a CUORE-0 bolometer. At low biases, the bolometer voltage (and thus its gain) increases linearly with bias current. However, at higher bias current, the joule heating across the NTD thermistor begins to cause electro-thermal feedback and heat the bolometer, lowering its gain. On the other hand, the baseline noise RMS tends to decrease at high bias current. Therefore the optimal working point is chosen to be where the signal-to-noise ratio, defined as the ratio of the amplitude of the reference pulse to the baseline noise RMS, is maximized. For the measurement shown in Figure 21 a bias current around 400 pA was chosen.

The time needed to perform the optimal working point measurement was dominated by the large number (~100) of five-second baseline measurements that were needed to properly evaluate the baseline noise RMS. To speed up the procedure, the measurement was performed in parallel on all the bolometers, and the baseline noise and pulser events were acquired every ~10 s and ~30 s, respectively. In this configuration the measurement lasted about 12 hours.

In their working configuration, the resistances of the CUORE-0 bolometers had an average value of 26 MΩ, with an RMS of 8 MΩ (see also Figure 23 and the discussion in Section 7.4). The bias currents were in the 100 – 400 pA range, with an average value of 250 pA and an RMS of 50 pA. The average CUORE-0 bolometer signal amplitude, defined as the ratio between the pre-gain pulse amplitude and the corresponding energy of the pulse, was 75.6 $\mu$V/MeV with an RMS of 32.3 $\mu$V/MeV among the 51 active channels. A typical 2615 keV calibration pulse recorded by a CUORE-0 bolometer is shown in Figure 22.

### 7.4 Technical runs

During the two data-taking campaigns of CUORE-0, we periodically performed technical measurements to monitor the performance of the bolometers.

One technical measurement was a scan with pulser events in the low energy range that was performed in each dataset to measure the trigger thresholds of the bolometers. Pulser events were generated every 40 s on each channel, ranging from 0 to 200 keV with steps of ~20 keV. Each of these measurements lasted for a few hours, during which about 50 pulses for each energy were acquired. From these measurements we estimate that the CUORE-0 energy thresholds, defined as the energy at which 90% of the pulses were detected by the standard signal trigger algorithm, were in the 30–120 keV range. The average threshold was ~70 keV.

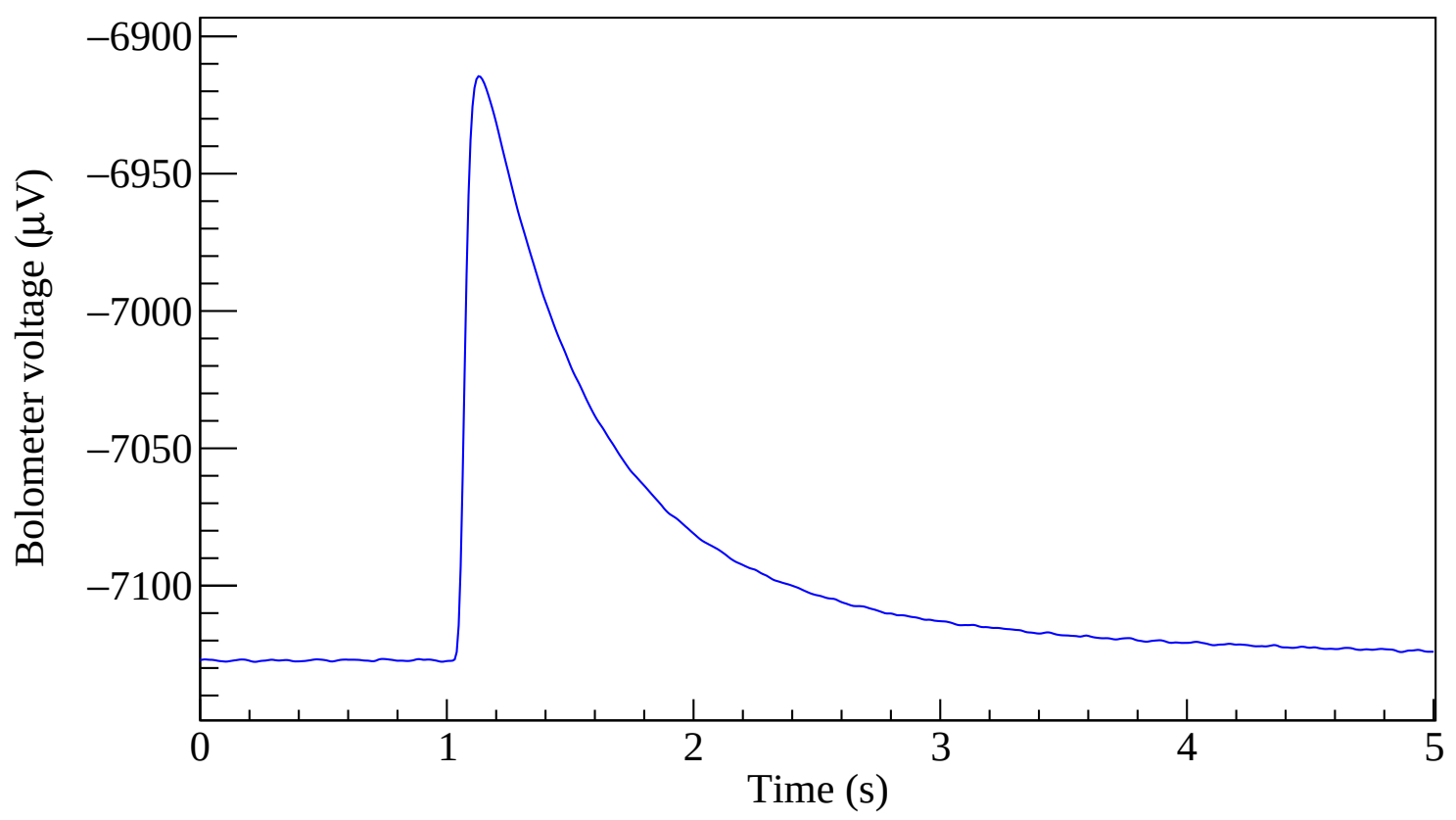

**Figure 22**. A pulse detected by a CUORE-0 bolometer, generated by an energy release of about 2615 keV during a calibration run. The voltage reported is referred to the ends of the NTD sensor, which was polarized with a current of ~250 pA.

A second measurement, performed daily, was a measurement of the detector resistances at their working point to track the detector stability over time. The resistance was obtained as the ratio of the bolometer voltage to the bias current, where the bolometer voltage was evaluated with the procedure described in Section 7.3. These measurements were only performed for the previously determined optimal value of the bias current and therefore took less time that the complete working point search described in Section 7.3, lasting about 10 minutes. Figure 23 shows an example of this measurement, repeated throughout a full dataset. At their working points, the resistances of the CUORE-0 bolometers were within a factor of 3 of each other, and they remained stable to within ~3% over the month-long dataset.

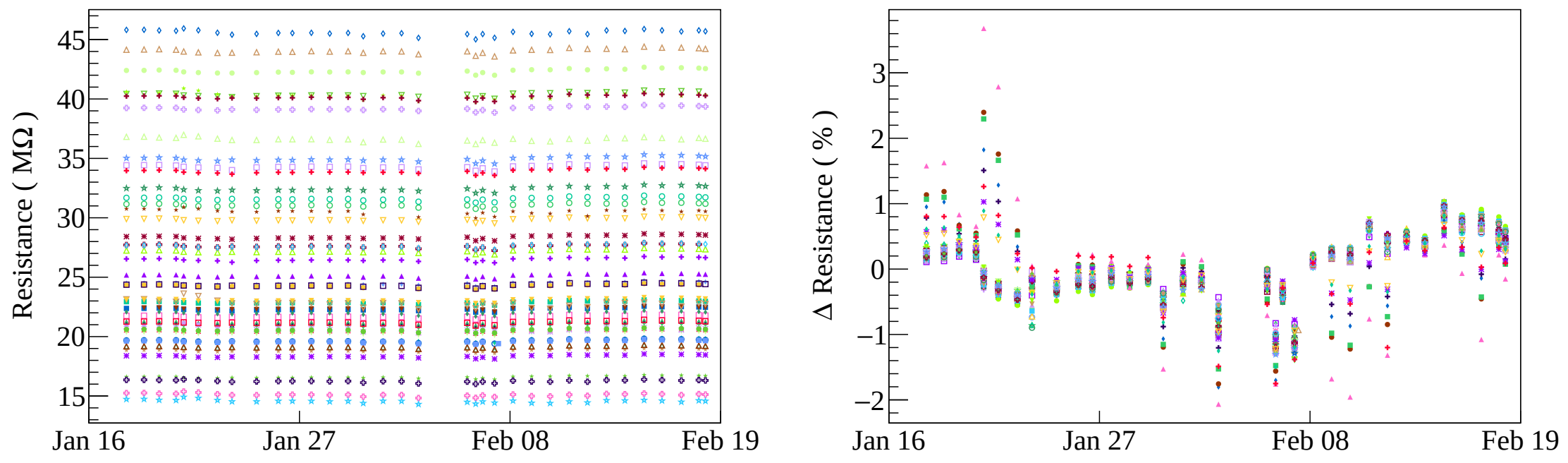

**Figure 23**. Left: Values of the CUORE-0 bolometer resistances at the working point measured daily over a period of one month in 2014. Right: Percentage variation of the same resistance values over the same time period. In both plots different marker colors and styles identify different bolometers.

### 7.5 Detector calibration and energy resolution

The detector energy resolution is a crucial parameter in any experiment searching for $0\nu\beta\beta$ decay, as it determines the detector's power to discriminate a signal peak from the background. The energy resolution of the CUORE-0 detector was evaluated using calibration data collected while the detector was exposed to thoriated tungsten wire sources. To calibrate two wires were inserted between the OVC of the cryostat and the external lead shield on opposite sides of the cryostat (as shown in Figure 15); each bolometer was then calibrated using the known $\gamma$ lines of the $^{232}$Th decay chain. The calibration energy spectrum exhibited a prominent $\gamma$ peak at 2615 keV from the decay of thorium-chain daughter $^{208}$Tl, and the width of this photopeak in each $TeO_2$ bolometer was used to estimate the bolometer's energy resolution at 2527.5 keV, the $Q$-value for $0\nu\beta\beta$ decay of $^{130}$Te. The $^{208}$Tl photopeak is a natural choice for this role because it is the closest high-statistics signal to the region of interest around 2527.5 keV. Figure 24 shows the sum of all the calibration spectra of the 51 CUORE-0 active channels.

Figure 25 shows the distributions of the energy resolutions evaluated on the 2615 keV calibration peak in Cuoricino and CUORE-0, for each bolometer in each dataset. The effective mean of the FWHM values in CUORE-0, is 4.9 keV (with a corresponding RMS of 2.9 keV); in Cuoricino, this was 5.8 keV with a RMS of 2.1 keV. Thus, the CUORE goal of 5 keV FWHM in the ROI has been achieved in CUORE-0. More details on the analysis can be found in [5, 42].

### 7.6 Background rate

In the energy region between 2.7 and 3.9 MeV, the background is dominated by degraded $\alpha$ particles and the contribution to the background from $\beta/\gamma$ events is negligible. Anticoincidence spectra (from events which deposited energy only in one crystal) in and around this energy region are shown for both CUORE-0 and Cuoricino in Figure 26. A flat background of events extends from about 4 MeV down to and below the $0\nu\beta\beta$ ROI. We excluded the $^{190}$Pt peak at 3150 keV because it has been shown to be bulk contamination and it does not affect the $0\nu\beta\beta$ background. In CUORE-0 a rate of 0.016 ± 0.001 counts/(keV·kg·y) was measured for this flat background in the $\alpha$ region [5]. This background value is a factor of 7 smaller than in Cuoricino, which measured 0.110 ± 0.001 counts/(keV·kg·y) in the same region [15]. This $\alpha$ continuum was the major contribution to the Cuoricino background in the ROI, but formed only a subdominant component of the ROI background in CUORE-0.

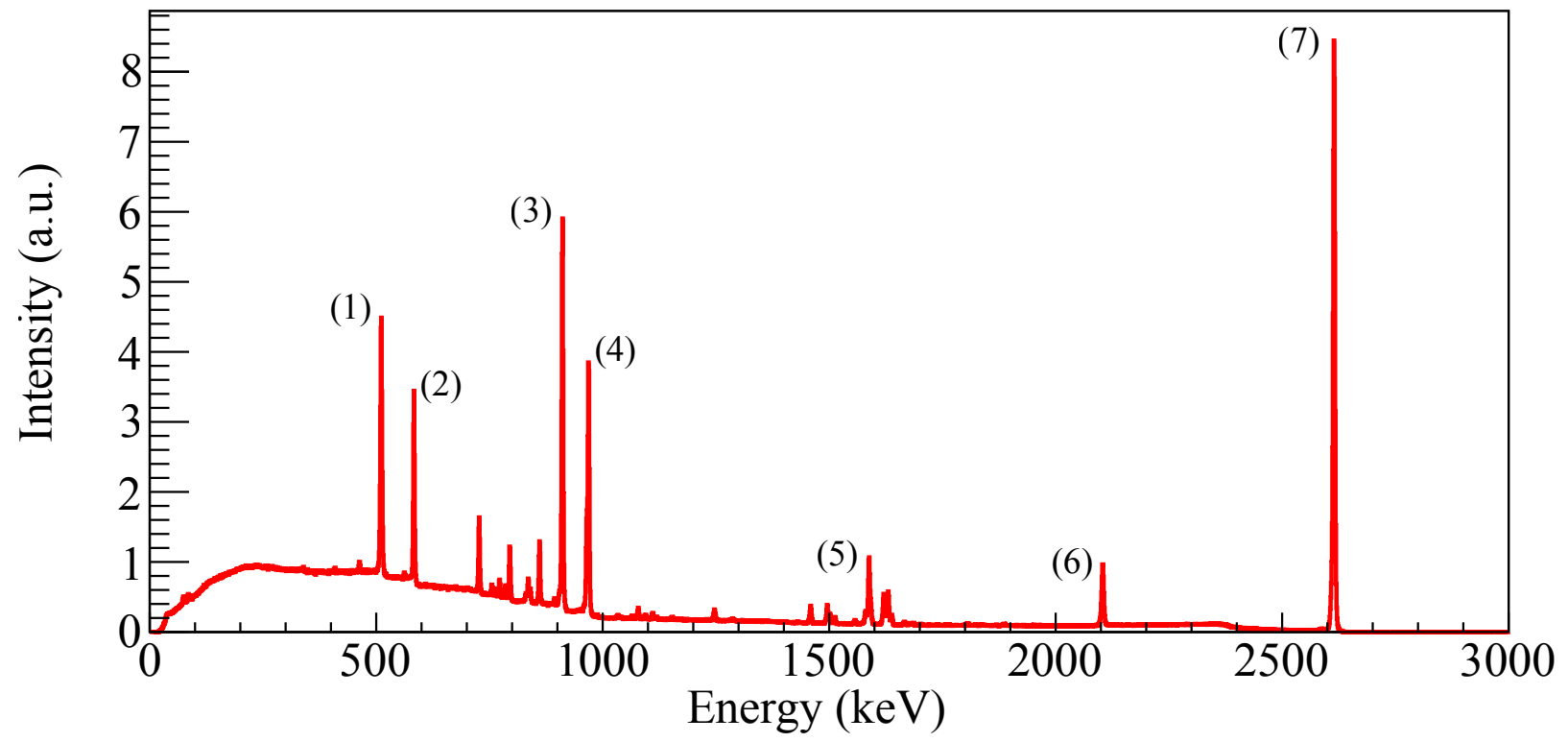

**Figure 24**. CUORE-0 calibration spectrum. The peaks are identified as $\gamma$ lines from the decay of nuclei in the $^{232}$Th decay chain: (1) 511 keV ($e^+e^-$ annihilation), (2) 583 keV ($^{208}$Tl), (3) 911 keV ($^{228}$Ac), (4) 965 keV and 969 keV ($^{228}$Ac), (5) 1588 keV ($^{228}$Ac), (6) 2104 keV ($^{208}$Tl with single 511 keV escape), (7) 2615 keV ($^{208}$Tl).

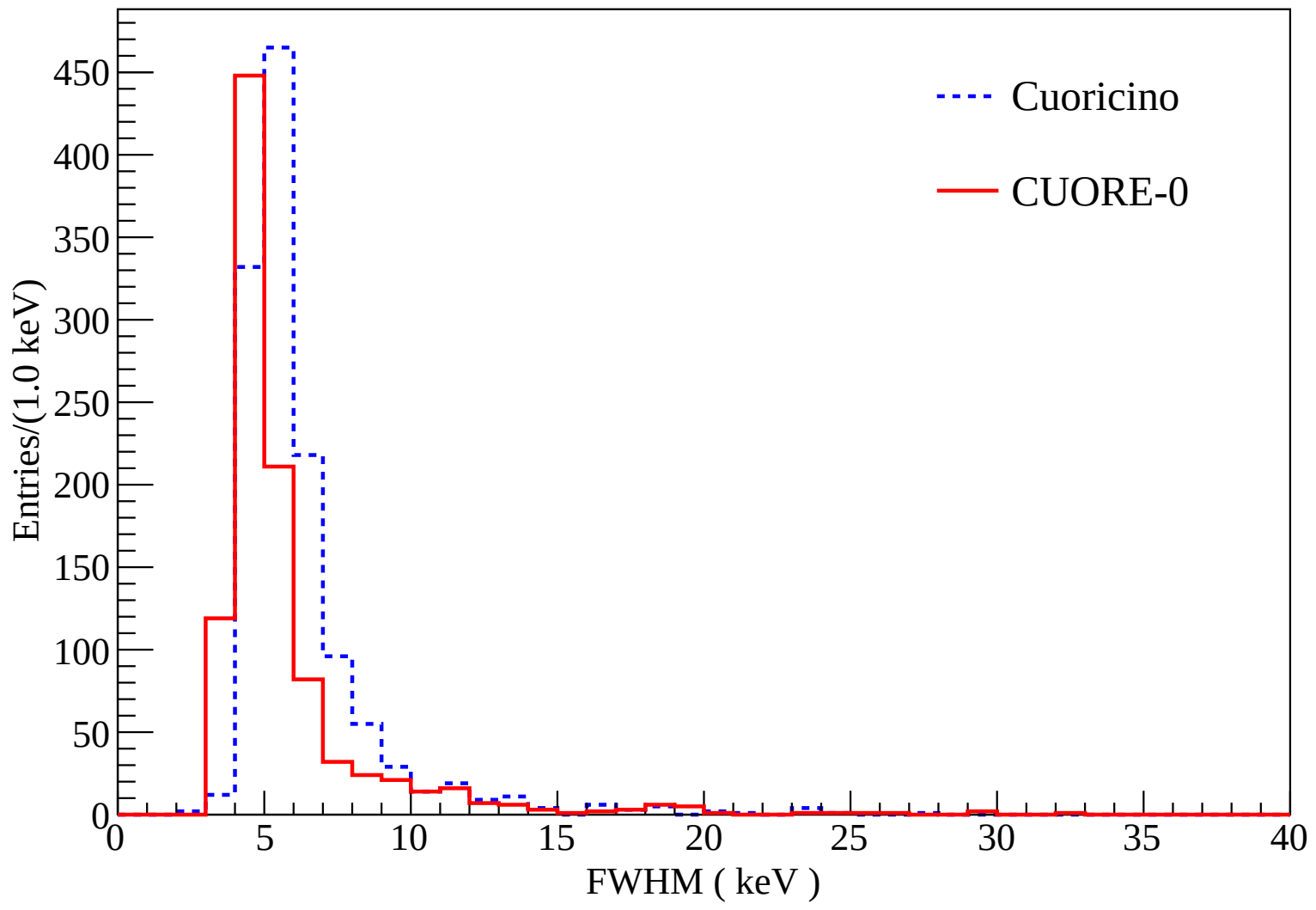

**Figure 25**. Comparison of the distribution of the energy resolutions (FWHM) measured in calibration for Cuoricino (blue dashed line) and CUORE-0 (red line).

In the ROI, the background rate measured by CUORE-0 was 0.058±0.004 (stat.)±0.002 (syst.) counts/(keV·kg·y)[5], which is an improvement of a factor of 3 from Cuoricino (0.169 ± 0.006 counts/(keV·kg·y) [15]). In this region of the energy spectrum, the background was comprised not only of degraded-$\alpha$ events, but also of $\beta/\gamma$ events produced by multi-Compton interactions of the 2615 keV $\gamma$ rays from $^{208}$Tl decay. According to Monte Carlo simulations, the source of this background was mostly located in the outer shields of the cryostat. Because CUORE-0 was operated in the same cryostat as Cuoricino, we did not expect to strongly suppress this background in the ROI. However, due to better material selection and to scaling effects, we expect a much smaller rate in CUORE. Therefore we expect the background in the ROI in CUORE will be mostly due to surface contamination by $\alpha$ emitters. By using the measured alpha background index in CUORE-0 as input to Monte Carlo simulations of CUORE, we conclude that the background goal of $10^{-2}$ counts/(keV·kg·y) is within reach. More details of the various background contributions are discussed in [64].

## 8 Conclusions

CUORE-0 served as a test of the new CUORE assembly line and achieved excellent results in terms of detector performance and radioactive background suppression. These results demonstrate that the detector design, assembly, and performance match the goals for CUORE. The rig-orous materials selection process and newly deployed surface cleaning procedures have proven effective in suppressing the background in the ROI and culminated in an $\alpha$ background index of $0.016 \pm 0.001$ counts/(keV·kg·y) in CUORE-0, which is a 7 fold improvement over Cuoricino. We found the bolometric performance to be excellent, with representative signal amplitude and detector energy resolution of 75 $\mu$V/MeV and 4.9 keV at 2.6 MeV, respectively. Furthermore, we have shown the new assembly line and procedures result in robust and reproducible detector characteristics; in particular the RMS of the bolometer effective temperature distribution in CUORE-0 was 2%, which is a factor of ~5 narrower than the same quantity in the predecessor experiment. Finally, CUORE-0 has extended the $0\nu\beta\beta$ physics program in $^{130}$Te. When combined with the data from Cuoricino we achieved a new limit on $0\nu\beta\beta$ decay of $^{130}$Te, $T_{1/2}^{0\nu\beta\beta} > 4.0 \times 10^{24}$ y (90% C.L.), which is the most stringent limit to date for this isotope [5].

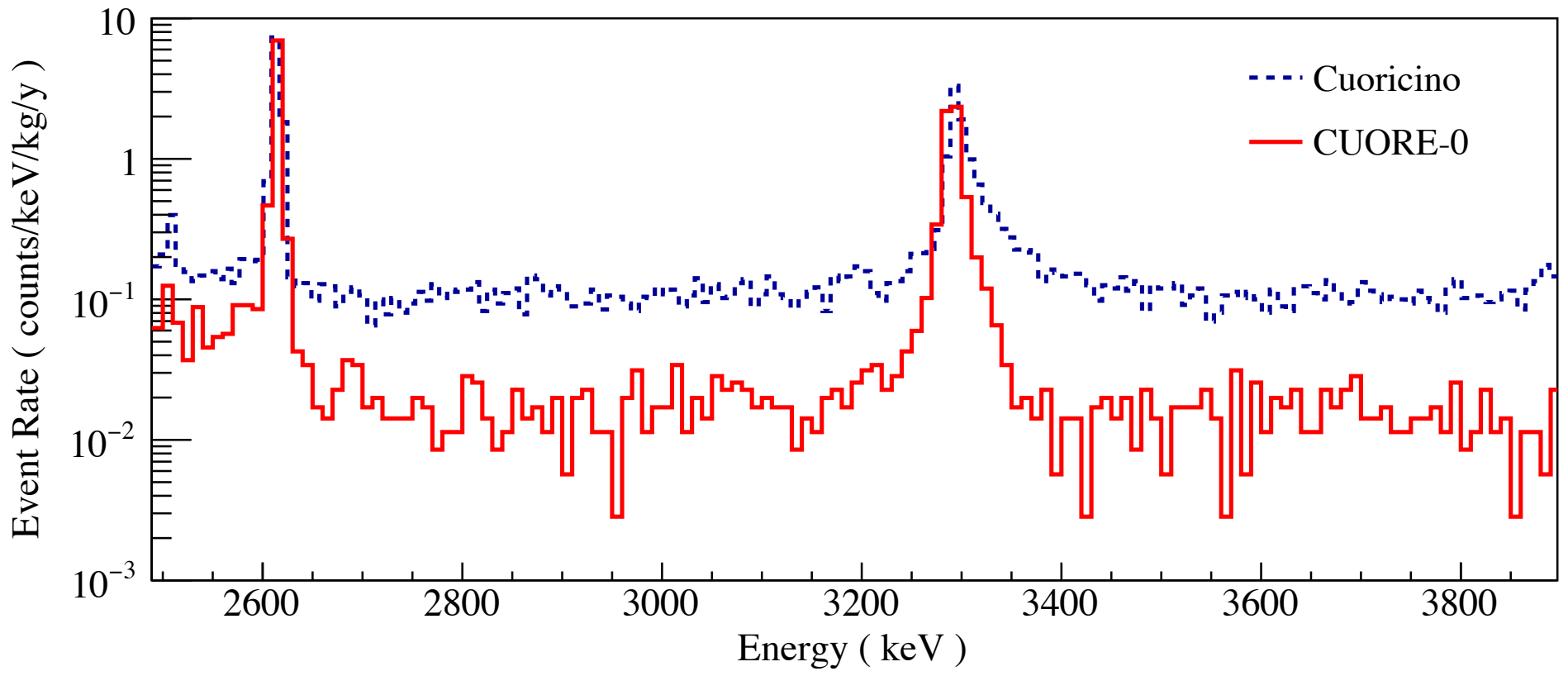

**Figure 26**. Comparison of the background index in the alpha region in Cuoricino (blue dashed line) and CUORE-0 (red solid line).

## Acknowledgments

The CUORE Collaboration thanks the directors and staff of the Laboratori Nazionali del Gran Sasso and the technical staff of our laboratories. This work was supported by the Istituto Nazionale di Fisica Nucleare (INFN); the National Science Foundation under Grant Nos. NSF-PHY-0605119, NSF-PHY-0500337, NSF-PHY-0855314, NSF-PHY-0902171, NSF-PHY-0969852, NSF-PHY-1307204, NSF-PHY-1314881, NSF-PHY-1401832, and NSF-PHY-1404205; the Alfred P. Sloan Foundation; the University of Wisconsin Foundation; and Yale University. This material is also based upon work supported by the US Department of Energy (DOE) Office of Science under Contract Nos. DE-AC02-05CH11231, DE-AC52-07NA27344, and DE-SC0012654; and by the DOE Office of Science, Office of Nuclear Physics under Contract Nos. DE-FG02-08ER41551 and DE-FG03-00ER41138. This research used resources of the National Energy Research Scientific Computing Center (NERSC).